\documentstyle[prb,multicol,eqsecnum,aps,epsf]{revtex}
\def\begineq
{\end{multicols}\widetext
  {\tiny           \noindent\begin{tabular}[t]{c|}
                   \parbox{0.48\hsize}{~} \\ \hline \end{tabular}}
   \vspace*{0.1cm}}
\def\endeq
{{\tiny\vspace*{0.1cm}\hspace*{\fill}\begin{tabular}[t]{|c}\hline
                    \parbox{0.475\hsize}{~} \\
                    \end{tabular}}\vspace*{0.cm}\begin{multicols}{2}
\columnseprule 0pt\narrowtext\noindent}
\def\d{d}
\def\lan{\left\langle}
\def\ran{\right\rangle}
\def\dl{{\frac{d}{dl}} \,}
\def\dx{dx}
\def\e{{\rm e}}
\def\virg{,}
\def\point{.}
\def\vf{v_{\rm F}}
\def\ef{\epsilon_{\rm F}}
\def\kf{k_{\rm F}}
\def\al{\alpha }
\def\para{\parallel }

\def\N{\hat{N}}
\def\tvp{\tilde{v}_{\sigma}}
\def\tvt{\tilde{v}_{\rho}}
\def\tvn{\tilde{v}_{\nu}}

\def\g2{g_{2}}

\def\tt{\tilde{t}}

\def\ggs{\buildrel\textstyle > \over {\hbox{\raise0.2ex\hbox{$\sim$}}}}
\def\lls{\buildrel\textstyle < \over {\hbox{\raise0.2ex\hbox{$\sim$}}}}
\def\gsim{\,\lower0.75ex\hbox{$\ggs$}\,}
\def\lsim{\,\lower0.75ex\hbox{$\lls$}\,}
\def\f{^{\rm F}}
\def\Krho  {K_{\rho}}
\def\Ksigma{K_{\sigma}}
\def\Kc   {K_{\rm C}}
\def\Ks   {K_{\rm S}}
\def\Knu  {K_{\nu}}
\def\gncp {G_{\nu + ,{\rm C} +}}
\def\gncm {G_{\nu + ,{\rm C} -}}
\def\gnc  {G_{\nu + ,{\rm C} p}}

\def\gnsp {G_{\nu + ,{\rm S} +}}
\def\gnsm {G_{\nu + ,{\rm S} -}}
\def\gns  {G_{\nu + ,{\rm S} p}}
\def\gnsd  {G_{\nu + ,{\rm S} p'}}

\def\gcs  {G_{{\rm C} p ,{\rm S} p'}}
\def\gcsp {G_{{\rm C} p ,{\rm S} +}}
\def\gcsm {G_{{\rm C} p ,{\rm S} -}}

\def\gcps {G_{{\rm C} + ,{\rm S} p}}
\def\gcms {G_{{\rm C} - ,{\rm S} p}}
\def\gcpsp{G_{{\rm C} + ,{\rm S} +}}
\def\gcpsm{G_{{\rm C} + ,{\rm S} -}}

\def\gcmsp{G_{{\rm C} - ,{\rm S} +}}
\def\gcmsm{G_{{\rm C} - ,{\rm S} -}}

\def\grc  {G_{\rho +, {\rm C} p}}
\def\grcp {G_{\rho +, {\rm C} +}}
\def\grcm {G_{\rho +, {\rm C} -}}
\def\grs  {G_{\rho +, {\rm S} p}}
\def\grsd {G_{\rho +, {\rm S} p'}}
\def\grsp {G_{\rho +, {\rm S} +}}
\def\grsm {G_{\rho +, {\rm S} -}}
\def\gpc  {G_{\sigma +, {\rm C} p}}
\def\gpcp {G_{\sigma +, {\rm C} +}}
\def\gpcm {G_{\sigma +, {\rm C} -}}
\def\gps  {G_{\sigma +, {\rm S} p}}
\def\gpsd {G_{\sigma +, {\rm S} p'}}
\def\gpsp {G_{\sigma +, {\rm S} +}}
\def\gpsm {G_{\sigma +, {\rm S} -}}

\def\gtb{\tilde{g}_1}
\def\gtf{\tilde{g}_2}
\def\gt{\tilde{g}}
\def\gtu{\tilde{g}_3}
\def\gtuc{\tilde{g}_{3 {\rm c}}}
\def\jo #1#2#3#4{#1 {\bf #2}, #4  (#3)} 

\def\PRB{Phys.\ Rev.\ B}
\def\PRL{Phys.\ Rev.\ Lett.}

\def\JPF{J.\ Phys.\ France}
\def\JPIF{J.\ Phys.\ I\ France}

\def\JPSJ{J.\ Phys.\ Soc.\ Jpn.}

\def\RMP{Rev.\ Mod.\ Phys.}
\def\PTP{Prog.\ Theor.\ Phys.}

\def\ADV{Adv.\ Phys.}

\def\SL{JETP\ Lett.}
\def\PHB{Physica B}

\def\PL{Phys.\ Lett.}
\def\JLP{J.\ Low Temp.\ Phys.}
\def\IJMPB{Int.\ J.\ Mod.\ Phys.\ B}

\def\SM{Synth.\ Met.}

\def\MCLC{Mol.\ Cryst.\ Liq.\ Cryst.}
\def\PS{Physica\ Scripta}
\begin{document}
\small
\draft

\title{
Confinement-deconfinement transition 
 in two-coupled chains  with umklapp scattering
}
\author{
M.  Tsuchiizu
and 
Y.  Suzumura
\\
}
\address
{
Department of Physics, Nagoya University, Nagoya 464-8602, Japan \\
}

\date{Received 4 November 1998;
To be published in Phys. Rev. B
}
\maketitle

\begin{abstract}
\footnotesize
A role of  umklapp scattering has been examined for 
  two-coupled  chains with both forward and backward scatterings
  by applying  renormalization group method to 
  bosonized Hamiltonian.  
It has been found that   
  a state with relevant interchain hopping changes into
  a state with irrelevant (confined) one
    when the magnitude of umklapp scattering  
  becomes larger than that of interchain hopping. 
 Critical value of umklapp scattering for 
 such a confinement-deconfinement transition 
  is calculated   as the function of 
  interchain hopping and intrachain interactions.    
 A crossover  from one-dimensional regime 
  into that of coupled chains 
  is also shown with decreasing temperature. 
\end{abstract}

\vspace*{-0.2cm}

\sloppy

\begin{multicols}{2}
\columnseprule 0pt
\narrowtext
\addtolength{\baselineskip}{0.2mm}

\section{Introduction}
Quasi-one-dimensional organic conductors, 
  (TMTSF)$_2$X and  (TMTTF)$_2$X salts, 
  exhibit instabilities  toward 
spin Peierls state, spin density wave (SDW)   state and 
 superconducting (SC) state, where the phase diagram 
 has been  displayed 
 on the plane of effective pressures and temperature.
\cite{Jerome,Bechgaard}
  The interplay of low-dimensionality and repulsive interaction 
  is important  for the SDW state which indicates   
  one-dimensional fluctuations.\cite{Gruner,Bourbonnais_JPF} 
There are also some evidences for dimensional crossover. \cite{Moser}

Crystal  structure shows quarter-filling for 
 conduction electrons  
  but  the  existence of   dimerization  leads to  
   half-filled band.\cite{Emery} 
A crossover from half-filled band to quarter-filled one 
  has been found by decreasing  dimerization  
  under effective pressure, i.e., 
  the variation of anions, X.  
  Electronic properties, which suggest  a role of  
  the dimerization,   have been reported recently 
  at temperatures just  above the SDW state.
  \cite{Gruner_ICSM,Gruner_prepri}
Optical experiments on a series of above materials, which 
  have  different values of interchain electron transfer energy,   
  show a correlation gap due to umklapp scattering and a crossover  
  from metallic state to insulating state with increasing the
  anisotropy. 
An insulator to metal transition followed by the deconfinement of 
  interchain hopping  has been observed when the interchain transfer
  energy exceeds a critical value  
  with a magnitude of the order of the gap.

Theoretical studies of these conductors have been explored  
  by use of quasi-one-dimensional model 
  consisting of an array of chains coupled by interchain hopping.
For repulsive intrachain interaction and incommensurate band,   
 the transverse  hopping   
  is always relevant for the weak interaction  
  \cite{Yakovenko} 
  but there is a reduction of transverse hopping  
  by one-dimensional fluctuation.\cite{Bourbonnais}
 Two-coupled chains is a basic  model for quasi-one-dimensional system 
  since 
  both one-dimensional fluctuation and transverse hopping 
 can be studied on the same footing. 
  In  a  Tomonaga-Luttinger 
  model with 
  only  forward    scattering,    
 the dominant state remains  the same as that of one-dimensional system but  
  the degeneracy of  in-phase and out-of-phase pairings is removed.
\cite{Nersesyan,Fabrizio,Finkelstein,Yoshioka_JLTP} 
 When backward scattering is added, 
 the  phase diagram  becomes  quite different from that 
 of a single chain. 
 In a Hubbard model with  repulsive interaction 
  and without umklapp scattering, 
  the ground state of two-coupled  chains is 
   the $d$-wave like  SC state\cite{Fabrizio,Rice,Schulz,Balents} 
 although that  of a single chain is 
  the SDW state.\cite{Solyom} 
   The effect of interchain hopping  is 
  much strong compared with  the intrachain interaction 
 since  the transverse  hopping  is   
   relevant  except for extremely large  intrachain 
  interaction.\cite{Fabrizio}
However  intrachain interaction becomes important 
  as well as the interchain hopping  
  for the case of the spin anisotropic backward scattering 
  where a spin gap induced in a single chain 
  leads to a competition between SDW state and SC state.\cite{T2} 
 
Recently confinement which denotes 
  incoherence of single particle hopping between 
  Luttinger liquids has been maintained,\cite{Anderson,Clarke} 
  where there is no coherence
  of hopping and then no split Fermi surface
  below a critical value of single particle hopping.
 The confinement has been argued for the metallic state of 
  organic conductor (TMTSF)$_2$X under a magnetic field,  
  which is close to coherence-incoherence transition.\cite{Strong} 
 The  role of umklapp scattering, which leads to the relevance 
 and the irrelevance of the correlation gap, has been 
examined for organic conductors.\cite{Bourbonnais_U}
In terms of a Mott gap, the irrelevance  of single
  particle hopping has been discussed in quasi-one-dimensional system.
  \cite{Giamarchi_physica,Giamarchi} 
A confinement 
 has been demonstrated   in two-coupled chains 
 with half-filled band\cite{Suzumura} 
  in order to understand a crossover from metallic state 
 to insulating state,  
 which has been found   at temperatures 
 just above SDW state of organic conductors.\cite{Gruner_prepri}
The weakly coupled half-filled chains with infinite numbers 
 have been also studied 
 by a perturbative renormalization group approach.
\cite{Bourbonnais_U,Kishine}

In the present paper, 
  such a deconfinement-confinement transition
  due to  umklapp scattering 
  is studied 
in details for  two-coupled chains with  half-filled band 
  by developing the previous work.\cite{Suzumura} 
In section II, formulation is given in terms of bosonized phase Hamiltonian.  
Renormalization group equations are derived  
  for coupling constants and response functions. 
In section III, the critical value for confinement is  calculated.  
A crossover at finite temperatures is also shown.
In section IV, we discuss  
  the validity of our present calculation and examine 
  an effect of forward scattering within the same branch.

\vspace{-0.5cm}
\section{Formulation} 

We consider  two-coupled chains given by 
\end{multicols}\widetext
\begin{eqnarray}            \label{H0}
{\cal H}
&=& \sum_{k,p,\sigma,i} \, 
    \epsilon_{k,p} a_{k,p,\sigma,i}^\dagger a_{k,p,\sigma,i}
{} -  t \sum_{k,p,\sigma} 
    \left[ 
    a_{k,p,\sigma,1}^\dagger a_{k,p,\sigma,2} + {\rm h.c.}
    \right] \nonumber \\ && \nonumber \\
& & {}+\frac{g_{1}}{2L} \sum_{p,\sigma,\sigma ',i} \sum_{k_1,k_2,q}
 a_{k_1,p,\sigma,i}^\dagger a_{k_2,-p,\sigma',i}^\dagger
    a_{k_2+2p\kf +q,p,\sigma',i} a_{k_1-2p\kf -q,-p,\sigma,i} 
   \nonumber \\&& \nonumber \\
& & {}+\frac{g_{2}}{2L} \sum_{p,\sigma,\sigma ',i} \sum_{k_1,k_2,q}
    a_{k_1,p,\sigma,i}^\dagger a_{k_2,-p,\sigma ',i}^\dagger
    a_{k_2+q,-p,\sigma ',i} a_{k_1-q,p,\sigma,i} \nonumber \\
&& \nonumber \\
& & {}+\frac{g_{3}}{2L} \sum_{p,\sigma,i} \sum_{k_1,k_2,q}
    a_{k_1,p,\sigma,i}^\dagger a_{k_2,p,-\sigma,i}^\dagger
    a_{k_2-2p\kf+q,-p,-\sigma,i} a_{k_1-2p\kf-q,-p,\sigma,i}  
\virg
\end{eqnarray}
\endeq
 where $t$ is the interchain hoping energy. 
 The quantity  $a_{k,p,\sigma,i}^\dagger $
 denotes a  creation  operator for  the electron
 with momentum $k$, spin 
  $\sigma(=\uparrow,\downarrow \,\,{\rm or}\,\, +,-)$ and chain index  $i(=1,2)$.  
  The symbol  $p=+(-)$ represents the right-going (left-going) state. 
In Eq.~(\ref{H0}), $\epsilon_{k,p}(=\vf (pk-k_{\rm F}))$ is
  the linearized kinetic  energy with Fermi velocity $\vf$ and 
   Fermi momentum $\kf$. 
  Quantities $g_2$,  $g_1$  and $g_3$ are 
 coupling constants of 
  intrachain interactions for     
 forward scattering, backward scattering and umklapp scattering, 
respectively. 

The diagonalization of 
 the first and the second terms in Eq.~(\ref{H0}) is performed   
 by making use of  an unitary transformation, 
$c_{k,p,\sigma,\mu} 
= ( - \mu a_{k,p,\sigma,1} + a_{k,p,\sigma,2} )/\sqrt{2} $
 with $\mu=\pm$.  
  After the bosonization of electrons 
 around the new Fermi point, 
 $k_{{\rm F}\mu} \equiv \kf - \mu t/\vf$, 
 we define the phase variables, 
  $\theta_{\rho +}$ and  $\theta_{\sigma +}$
($\theta_{{\rm C} +}$ and $\theta_{{\rm S}+}$),  
  which   express    fluctuations 
 of the total (transverse) charge density and 
  spin density.\cite{Yoshioka_JLTP} 
 They are given by  
\begin{eqnarray}
\theta_{\rho \pm} (x)   \label{eq:phase_rho}
         &=& \frac{1}{\sqrt{2}} \sum_{q\neq 0} \frac{\pi i}{qL}
             \e ^{-\frac{\al}{2}|q|-iqx} \sum_{k,\sigma,\mu}
\nonumber \\ && \nonumber \\ &&\times
    \left(  c_{k+q,+,\sigma,\mu}^\dagger c_{k,+,\sigma,\mu}
        \pm c_{k+q,-,\sigma,\mu}^\dagger c_{k,-,\sigma,\mu}
    \right), \nonumber \\&&
\\  && \nonumber \\
\theta_{\sigma \pm} (x)
         &=& \frac{1}{\sqrt{2}} \sum_{q\neq 0} \frac{\pi i}{qL}
             \e ^{-\frac{\al}{2}|q|-iqx} \sum_{k,\sigma,\mu} \sigma
\nonumber \\ && \nonumber \\ &&\times
    \left(  c_{k+q,+,\sigma,\mu}^\dagger c_{k,+,\sigma,\mu}
        \pm c_{k+q,-,\sigma,\mu}^\dagger c_{k,-,\sigma,\mu}
    \right), \nonumber \\&&
\\ && \nonumber \\
\theta_{{\rm C} \pm} (x)
         &=& \frac{1}{\sqrt{2}} \sum_{q\neq 0} \frac{\pi i}{qL}
             \e ^{-\frac{\al}{2}|q|-iqx} \sum_{k,\sigma,\mu} \mu
\nonumber \\ && \nonumber \\ &&\times
    \left(  c_{k+q,+,\sigma,\mu}^\dagger c_{k,+,\sigma,\mu}
        \pm c_{k+q,-,\sigma,\mu}^\dagger c_{k,-,\sigma,\mu}
    \right), \nonumber \\&&
\\
\theta_{{\rm S}\pm} (x)  \label{eq:phase_S}
         &=& \frac{1}{\sqrt{2}} \sum_{q\neq 0} \frac{\pi i}{qL}
             \e ^{-\frac{\al}{2}|q|-iqx} \sum_{k,\sigma,\mu}
             \sigma \mu 
\nonumber \\ && \nonumber \\ &&\times
    \left(  c_{k+q,+,\sigma,\mu}^\dagger c_{k,+,\sigma,\mu}
        \pm c_{k+q,-,\sigma,\mu}^\dagger c_{k,-,\sigma,\mu}
    \right). \nonumber \\&&
\end{eqnarray}
 There is a  commutation
  relation that 
$  [\theta_{\nu +}(x),\theta_{\nu' -}(x')]
 = i \pi \delta_{\nu, \nu'}\,{\rm sgn}(x-x')$ 
 where the suffix $-$ denotes the  canonically conjugate variable. 
 In terms of these phase variables, the field operator is  expressed as 
\begin{eqnarray}
\psi_{p,\sigma,\mu}(x) & = & 
 L^{-1/2}  \sum_k \e^{ikx} c_{k,p,\sigma,\mu} 
\nonumber \\ 
&=&
 \frac{1}{\sqrt{2\pi \al} }
\exp \left( ipk_{{\rm F}\mu}x 
 + i\Theta _{p,\sigma,\mu} \right) 
   \exp \left( i\pi \Xi_{p,\sigma,\mu} \right),
\nonumber \\ 
\label{eqn:field} \\ && \nonumber \\
\Theta _{p,\sigma,\mu}
 & = &  
\frac{1}{2\sqrt{2}}
  [
     p \theta_{\rho +} + \theta_{\rho -} 
   + \sigma ( p \theta_{\sigma +} + \theta_{\sigma -} )
\nonumber \\  \nonumber \\ && {}
   + \mu    ( p \theta_{{\rm C}+} + \theta_{{\rm C}-} )
   + \sigma \mu ( p \theta_{{\rm S}+} + \theta_{{\rm S}-} )
                      ],
\end{eqnarray}
 where  $\al$ is of the order of the lattice constant.  
 The phase factor, $\pi \Xi_{p,\sigma,\mu}$, 
  in Eq.~(\ref{eqn:field}), 
 which is introduced for  the anticommutation relation,
   is taken  as 
\begin{eqnarray}
\Xi_{2n+j} &=& \N_1 + \cdots +\N_{2n} 
    + \frac{(-1)^{j+1}}{2} (\N_{2n+1} + \N_{2n+2})   \virg
\nonumber \\
\end{eqnarray}
where  $j$=1,2 and $n=0,1,2,3$.    
 The quantity, $\N_{i}$, denotes  number operator, 
and  the suffix $i$ is related to  
 ($p$, $\sigma$, $\mu$) as, 
 $(+,+,+)=1,(+,-,+)=2,(+,+,-)=3,
(+,-,-)=4,(-,+,+)=5,(-,-,+)=6,(-,+,-)=7$ 
 and $(-,-,-)=8$, respectively.
 Such a choice of   $\Xi_{p,\sigma,\mu}$ 
 conserves   a sign of interactions, which are represented 
 by phase operators.  
 In terms of these operators, 
  Eq.~(\ref{H0}) is rewritten as (Appendix A)
\begineq
\begin{eqnarray} \label{eqn:H}
{\cal H} &=& 
\sum_{\nu = \rho,\sigma,{\rm C},{\rm S}} \frac{v_\nu}{4\pi} \int \hspace{-1mm} \dx
 \left\{
   \frac{1}{K_\nu} \left(\partial \theta_{\nu +} \right)^2
         +  K_\nu  \left(\partial \theta_{\nu -} \right)^2
 \right\}
\nonumber \\   && \nonumber \\
&&{}+
    \frac{g_\rho}{4\pi^2 \alpha^2} \int \hspace{-1mm} \dx 
    \biggl\{ 
        \cos \biggl( \sqrt{2}\theta_{{\rm C}+} - \frac{4t}{\vf}x\biggr) 
      + \cos \sqrt{2} \theta_{{\rm C}-}  
    \biggr\}
    \left\{
        \cos \sqrt{2} \theta_{{\rm S} +} - \cos \sqrt{2} \theta_{{\rm S} -}
    \right\} \nonumber \\&& \nonumber \\
&&{}+
    \frac{g_\sigma}{4\pi^2 \alpha^2} \hspace{-1mm} \int \hspace{-1mm} \dx
    \biggl\{ 
        \cos \biggl( \sqrt{2}\theta_{{\rm C} +} -\frac{4t}{\vf}x\biggr) 
      - \cos \sqrt{2} \theta_{{\rm C} -}  
    \biggr\} 
    \left\{
        \cos \sqrt{2} \theta_{{\rm S} +} 
             + \cos \sqrt{2} \theta_{{\rm S} -}
    \right\} \nonumber \\ && \nonumber \\
&&{}+
    \frac{g_{1}}{2\pi^2 \alpha^2}  \int \hspace{-1mm}\dx
    \cos \sqrt{2} \theta_{\sigma +} 
    \biggl\{
        \cos \biggl( \sqrt{2}\theta_{{\rm C} +} - \frac{4t}{\vf}x\biggr)
      - \cos \sqrt{2} \theta_{{\rm C} -} 
      - \cos \sqrt{2} \theta_{{\rm S} +} 
      - \cos \sqrt{2} \theta_{{\rm S} -} 
    \biggr\}            \nonumber \\&& \nonumber \\
&&{}+
    \frac{g_{3}}{2\pi^2 \alpha^2} \int \hspace{-1mm}\dx
    \cos \sqrt{2} \theta_{\rho +} 
    \biggl\{
        \cos \biggl( \sqrt{2}\theta_{{\rm C}+} - \frac{4t}{\vf}x
\biggr)
      + \cos \sqrt{2} \theta_{{\rm C} -} 
      - \cos \sqrt{2} \theta_{{\rm S} +} 
      + \cos \sqrt{2} \theta_{{\rm S} -} 
    \biggr\}  \virg
\label{phase_Hamiltonian}
\end{eqnarray}
\endeq
where
$v_\nu=\vf \sqrt{1-\{g_\nu/2\pi\vf\}^2}$,  
$K_\nu = [ \{1-g_\nu/2\pi\vf\}
      /\{1+ g_\nu/2\pi\vf\} ]^{1/2}$,
$g_\rho = 2 g_2 -g_1$, $g_\sigma = -g_1$
 and 
$g_{\rm C } = g_{\rm S} = 0$. 

 We make use of renormalization group method  for
  response functions,
\cite{Giamarchi_JPF,Giamarchi_PRB,Tsuchiizu}
 which are assumed to be invariant for scaling, 
   $\alpha \to \alpha '=\alpha \e^{\d l}$.   
We express the nonlinear terms in Eq.~(\ref{phase_Hamiltonian})
  as $g_{\nu p, \nu' p'}/(2 \pi^2 \alpha^2) \int \dx \cos \sqrt{2}
  \bar{\theta}_{\nu p} \cos \sqrt{2}\bar{\theta}_{\nu' p'}$
  where $\sqrt{2}\bar{\theta}_{\nu p} =  \sqrt{2}\theta_{\nu p}
  -4tx/\vf$ for $\nu = {\rm C}$ and $p = +$ and 
   $\sqrt{2}\bar{\theta}_{\nu p} =  \sqrt{2}\theta_{\nu p}$ 
 otherwise.
Then the coupling constants are given by
  ${g_{{\rm C} + ,{\rm S} +}} = -{g_{{\rm C} - ,{\rm S} -}} 
         = (g_\rho + g_\sigma)/2$,
  ${g_{{\rm C} + ,{\rm S} -}} = -{g_{{\rm C} - ,{\rm S} +}} 
         = (g_\rho - g_\sigma)/2$,
  ${g_{\sigma +, {\rm C} +}} =-{g_{\sigma +, {\rm C} -}} 
         = -{g_{\sigma +, {\rm S} +}} =-{g_{\sigma +, {\rm S} -}} 
         = g_1$ and
  ${g_{\rho +, {\rm C} +}} = {g_{\rho +, {\rm C} -}} 
         = - {g_{\rho +, {\rm S} +}} = {g_{\rho +, {\rm S} -}} = g_3$.
Response functions defined by  
$R_{A}(x_1-x_2,\tau _1-\tau _2) 
\equiv 
\lan T_\tau O_A(x_1,\tau_1) \, O_A^\dagger (x_2,\tau_2) \ran $
 are evaluated  for SDW and SC states 
  where $\tau_j$ is the imaginary time and $O_A$ denotes 
 the order parameter.  
Then  renormalization group equations  are expressed as\cite{Suzumura} 
 (Appendix B) 
\begin{eqnarray}    
\dl K_\nu &=& 
- \frac{1}{2  \tilde{v}_\nu^2} K_\nu^2 \,  
   \biggl[
    \gncp^2  \,
    J_0 ( 4\tt ) 
\nonumber \\ && \nonumber \\ && {}
+  \gncm^2 + \gnsp^2 + \gnsm^2 
  \biggr] \,,
\label{K_theta}
\end{eqnarray}\begin{eqnarray}    
          \label{K:Theta}
\dl \Kc &=& 
  \frac{1}{2} \sum_{p=\pm} 
   \biggl[
     \left( - \Kc^2 \, J_0( 4\tt) \, \delta_{p,+} 
                                         + \delta_{p,-} \right)
\left\{\grc^2
\right. \nonumber \\ &&\nonumber \\ && \left. {} 
        + \gpc^2 + \gcsp^2 + \gcsm^2 \right\}
   \biggr] \,,
\label{K_thetatilde}
\end{eqnarray}\begin{eqnarray}    
\dl \Ks &=&
  \frac{1}{2} \sum_{p=\pm} 
   \biggl[
       \left( - \Ks^2 \, \delta_{p,+} + \delta_{p,-} \right)
       \left\{ \grs^2 
\right. \nonumber \\ && \nonumber \\ &&\left. {}
        + \gps^2  +  \gcps^2 \, J_0( 4\tt )  + \gcms^2
       \right\} 
   \biggr] \,,
\nonumber \\
\label{K_phitilde}
\end{eqnarray}\begin{eqnarray}    
          \label{G:Theta}
\dl \gnc &=& 
    \Bigl( 2 - K_\nu - \Kc^{p} \Bigr) \gnc
\nonumber \\ && \nonumber \\ &&{}
      - \gnsp \, \gcsp
   - \gnsm \, \gcsm 
 \virg  \end{eqnarray}\begin{eqnarray}    
          \label{G:Phi}
\dl \gns &=& 
    \Bigl( 2 - K_\nu - \Ks^{p} \Bigr) \gns
 - \gncp \,\gcps \, J_0( 4\tt ) 
\nonumber \\ && \nonumber \\ &&{}
   - \gncm \, \gcms
  \virg  \end{eqnarray}\begin{eqnarray}    
          \label{G:TL}
\dl \gcs &=& 
    \Bigl( 2 - \Kc^p - \Ks^{p'} \Bigr) \gcs
    - \frac{1}{\tilde{v}_\rho} \, \grc \, \grsd  
\nonumber \\ && \nonumber \\ &&{}
    - \frac{1}{\tilde{v}_\sigma} \, \gpc \, \gpsd 
 \virg \end{eqnarray}\begin{eqnarray}    
\dl \tt &=& 
    \tt 
 -  \frac{1}{4}  \Kc \Bigl( 
         \grcp^2  +  \gpcp^2
\nonumber \\ && \nonumber \\ && {}
              + \gcpsp^2 + \gcpsm^2 
           \Bigr)
         J_1(4\tt)
   \virg
\label{dt} 
\end{eqnarray}
 where
 $\tt(l) = t(l)/\ef$,  $\ef \equiv \vf \alpha^{-1}$, 
 $\tilde{v}_\nu = v_\nu /\vf$, $\nu = \rho$, $\sigma$ and
 $p$, $p' = \pm$.
In these equations, 
 the $l$-dependence is not written explicitly 
and
 $J_n$ $(n=0,1)$ is the $n$-th order Bessel function.\cite{correct}
 Initial conditions are given by 
$K_\nu (0) = K_\nu$,
$G_{\nu p, \nu' p'} (0) = g_{\nu p, \nu' p'}/2\pi\vf$
and 
$\tt (0) = t/\ef$.

 The second order renormalization group equations with respect to 
 all the coupling constants   
 are derived by expanding  as 
 $K_\nu ^{\pm 1} (l)= 1 \mp G_\nu (l) + \cdots$.
In case of $g_3 = 0$,\cite{T2}
these  equations become equal to those of Fabrizio,\cite{Fabrizio} 
 which satisfy  the   SU(2) symmetry with respect to spin rotation. 
 Although such a symmetry is satisfied only approximately 
  for Eqs.~(\ref{K_theta})-(\ref{K_phitilde}),  the 
  difference is very small within the present choice of parameters 
 as is shown later.  
The renormalization equations of 
  Eqs.~(\ref{K_theta})-(\ref{K_phitilde}) determine the fluctuations 
  of the total charge, total spin, transverse charge 
  and transverse spin density respectively.
Equation (\ref{G:TL}) corresponds to forward  scattering and 
  backward scattering with parallel spins. 
 In r.h.s. of these equations, 
 there are  bilinear terms with respect to 
 $G_{\nu p, \nu'  p'} (l)$, which appear 
 in the presence of  umklapp scattering 
  and/or backward scattering while  they are absent for 
  only forward scattering. 
 Equations (\ref{G:Theta}) and (\ref{G:Phi}) 
  with $\nu = \rho$ ($\nu = \sigma$)
  correspond to umklapp scattering
  (backward scattering with opposite spins).
It is found that there is a symmetry between  equations of 
  the total charge and those of the total spin, 
 i.e., the renormalization equations remain the same 
 for the  replacement given by  $K_\rho \leftrightarrow K_\sigma$, 
  $v_\rho \leftrightarrow v_\sigma$ and 
  $(G_{\rho +, {\rm C}+},\, G_{\rho +, {\rm C}-},\, 
     G_{\rho +, {\rm S}+},\, G_{\rho +, {\rm S}-}) \leftrightarrow 
  (G_{\sigma +, {\rm C}+},\, G_{\sigma +, {\rm C}-},\, 
     G_{\sigma +, {\rm S}+},\, G_{\sigma +, {\rm S}-}) $.
Equation  (\ref{dt}) is the scaling equation for the interchain hopping.
It is noted that these equations  with  $t=0$
 is reduced to those of  
  a single chain.\cite{Solyom}

We examine order parameters  for the possible states 
  in case of repulsive interaction.
In terms of phase variables, order parameters are expressed as
  (Appendix A)
\begin{eqnarray}
 O_{{\rm LSDW}_{\para,{\rm out}}}
 &=&         \sum_{\sigma} \sigma \,
       \{  \psi_{+,\sigma,1}^\dagger \, \psi_{-,\sigma,1} 
       - 
         \psi_{+,\sigma,2}^\dagger \, \psi_{-,\sigma,2} \}
\nonumber \\ && \nonumber \\
&=& \sum_{\sigma,\mu} \sigma \, 
         \psi_{+,\sigma,\mu}^\dagger \, \psi_{-,\sigma,-\mu}
\nonumber \\&& \nonumber \\
&\to&
        \e^{-i2\kf x}
         \displaystyle{\sum_{\sigma}}   
         \exp \left[ -i
           \bigl\{ \theta_{\rho +} + \sigma \theta_{\sigma +}
           \bigr\}/\sqrt{2}
              \right] \,
\nonumber \\ && \nonumber \\ && {} \times
         \cos \left[
         \left\{ \theta_{{\rm C} -} + \sigma \theta_{{\rm S} -}
       \right\}/\sqrt{2}
         \right]   
 \label{O_LSDW} 
 \virg \\ && \nonumber \\
 O_{{\rm TSDW}_{\para,{\rm out}}}
 &=&         \sum_{\sigma} 
       \{  \psi_{+,\sigma,1}^\dagger \, \psi_{-,-\sigma,1} 
       - 
         \psi_{+,\sigma,2}^\dagger \, \psi_{-,-\sigma,2} \}
\nonumber \\&& \nonumber \\
&=& \sum_{\sigma,\mu}
    \psi_{+,\sigma,\mu}^\dagger \, \psi_{-,-\sigma,-\mu}
\nonumber \\&& \nonumber \\
&\to&
      \e^{-i2\kf x}
         \displaystyle{\sum_{\sigma}} \sigma   
         \exp \left[ -i
           \bigl\{ \theta_{\rho +} + \sigma \theta_{\sigma -}
           \bigr\}/\sqrt{2}
              \right] \,
\nonumber \\ && \nonumber \\ && {} \times
         \sin \left[
         \left\{ \theta_{{\rm C} -} + \sigma \theta_{{\rm S}+} 
            \right\}/\sqrt{2}
         \right]   
 \label{O_TSDW} 
\virg \\ 
  O_{{\rm SS}_{\perp,{\rm in}}}
 &=&         \sum_{\sigma}   \sigma
   \{      \psi_{+,\sigma,1} \, \psi_{-,-\sigma,2} 
       + 
         \psi_{+,\sigma,2} \, \psi_{-,-\sigma,1}\}
\nonumber \\&& \nonumber \\
&=& \sum_{\sigma,\mu} \sigma \mu \,
         \psi_{+,\sigma,\mu} \, \psi_{-,-\sigma,\mu}
\nonumber \\&& \nonumber \\
&\to&
      \sum_{\sigma}  \sigma \,
          \exp \left[ i
               \bigl\{ \theta_{\rho -} + \sigma \theta_{\sigma +}  
               \bigr\} /\sqrt{2}
               \right] \,
\nonumber \\ && \nonumber \\ && {} \times
          \sin \left[
               \left\{ \theta_{{\rm C} -} + \sigma\theta_{{\rm S}+} 
               \right\}/
         \sqrt{2}
          \right] 
 \label{O_SS} 
                \point  
\end{eqnarray}
where   
$\psi_{p,\sigma,i} (x) = (1/\sqrt{L} ) 
 \sum_k \e^{ikx} a_{k,p,\sigma,i}$. 
 In Eqs.~(\ref{O_LSDW})-(\ref{O_SS}), 
 ${\rm LSDW}_{\para,{\rm out}}$  (${\rm TSDW}_{\para,{\rm out}}$) 
 denotes  longitudinal (transverse) SDW 
 with intrachain and out-of-phase pairing.   
The suffix, ${\rm SS}_{\perp,{\rm in}}$, 
 represents SCd state, i.e., the singlet SC state with interchain 
 and in-phase pairing.

The renormalization group technique is also applied to the calculation 
 of  response functions for the order parameters, 
  Eqs.~(\ref{O_LSDW})-(\ref{O_SS}). 
Normalized response functions are derived as (Appendix B)
\begineq
\begin{eqnarray}
\overline{R}_{{\rm LSDW}_{\para,{\rm out}}}(r)
&=& \exp \biggl[ \int_0^{\ln (r/\al)} \d l 
       \biggl\{ -\frac{1}{2} 
       \Bigl(  K_\rho (l) + K_\sigma (l) 
          + 1/K_{\rm C} (l) + 1/K_{\rm S} (l)
       \Bigr) 
\nonumber \\ && \hspace{6cm}
       - \gcmsm (l) - \gpcm (l) - \gpsm (l)
    \biggr\} \biggr]  \virg
\label{R-LSDW}
\\ && \nonumber \\
\overline{R}_{{\rm TSDW}_{\para,{\rm out}}}(r)
&=& \exp \biggl[ \int_0^{\ln (r/\al)} \d l 
       \biggl\{ -\frac{1}{2} 
       \Bigl(  K_\rho (l)  + 1/K_\sigma (l) 
          + 1/K_{\rm C} (l) + K_{\rm S} (l)
       \Bigr)
       + \gcmsp (l) 
    \biggr\} \biggr]    \label{R-TSDW}
\virg \\ && \nonumber \\
\overline{R}_{{\rm SS}_{\perp,{\rm in}}}(r)
&=& \exp \biggl[ \int_0^{\ln (r/\al)} \d l 
       \biggl\{ -\frac{1}{2} 
       \Bigl(  1/K_\rho (l) + K_\sigma (l) 
          + 1/K_{\rm C} (l) + K_{\rm S} (l)
       \Bigr) 
\nonumber \\ && \hspace{6cm}
       + \gcmsp (l) - \gpcm (l) + \gpsp (l) 
    \biggr\} \biggr]  \virg \label{R-SS}
\end{eqnarray}
\endeq
where $r = [x^2 + (\vf \tau)^2]^{1/2}$ 
 and the quantities $K_\nu (l)$ 
  ($\nu = \rho$, $\sigma$, ${\rm C}$ and ${\rm S}$) and 
  $G_{\nu p, \nu' p'} (l)$ ($p$, $p' = \pm$) are calculated from 
  Eqs.~(\ref{K_theta})-(\ref{G:TL}).
In these  equations, the renormalization
  of the velocity \cite{Giamarchi_JPF}
   has been discarded in a way similar 
to  the spinless case.\cite{Tsuchiizu}

\section{ Confinement vs. Deconfinement}

We examine confinement-deconfinement transition 
  by calculating the renormalization group equations 
 for interactions of both Hubbard model and
    general model with  $g_1 \neq g_2 $.
The scaling quantity $l \, (= \ln r/\alpha)$ is related to 
 energy $\omega$ and/or temperature $T$ by the relation that $l=$
  $\ln (\ef/\omega) = \ln (\ef/T)$. 
 Numerical calculation is performed by use of 
 normalized quantities 
  $\gt_j \equiv g_j/(2\pi\vf)$ for $j=1 \sim 3$.
In Fig.~1(a), quantities 
$\tt (l)$ and $1/\Kc(l)$  as a function of $l$ 
  are shown 
  by  solid  curves and dashed curves, respectively 
  with the fixed $\gtu=$ 0.05, $\gtuc$ (=0.119) and 0.3   
  where $t/\ef =0.1$ and $\gt_1 = \gt_2  = 0.4$.  
  Both mutual interactions and   umklapp scattering
   suppress the increase of  $t(l)$ as is seen from Eq.~(\ref{dt}).  
In case of   $\gtu$ =0.05,  
   $\tt (l)$ (solid curve (1))   increases rapidly.  
Such a behavior of $\tt(l)$ denotes  the deconfinement 
  of the transverse hopping. 
The corresponding    
   $1/\Kc(l)$ shown by  dashed curve (I)  
  decreases  monotonically to zero indicating 
  a formation of the transverse charge gap.   
  In the present case,  some of $G_{\nu p, \nu' p'}(l)$ diverge 
 at finite $l$  and then solutions stop 
 due to the second order  renormalization group equations. 
  It is expected that 
   the calculation with  third order equations 
 gives  finite value of $G_{\nu p, \nu' p'}(l)$ 
   for  all value of $l$.\cite{Fabrizio} 
 A noticeable difference appears for large value 
 of the umklapp scattering as is  shown 
  for $\gt_3 =  0.3$ (curves  (3) and (III)). 
 With increasing $l$,  $\tt(l)$ (curve (3)) 
takes a maximum and 
 reduces to zero     and   $1/K_{\rm C}(l)$ (curve (III)) 
 remains finite even   at the limiting value of $l$.
Such a behavior of $\tt(l)$  indicates   
  the absence of 
  interchain hopping  which leads to confinement of electrons
  within a single chain.   
There is no transverse charge gap 
  due to finite $K_{\rm C}(l)$, where   
 the oscillatory behavior   comes from   the Bessel functions 
  in 
\begin{figure}[htb]
\begin{flushleft}
\leavevmode
\epsfxsize=3.in\epsfbox{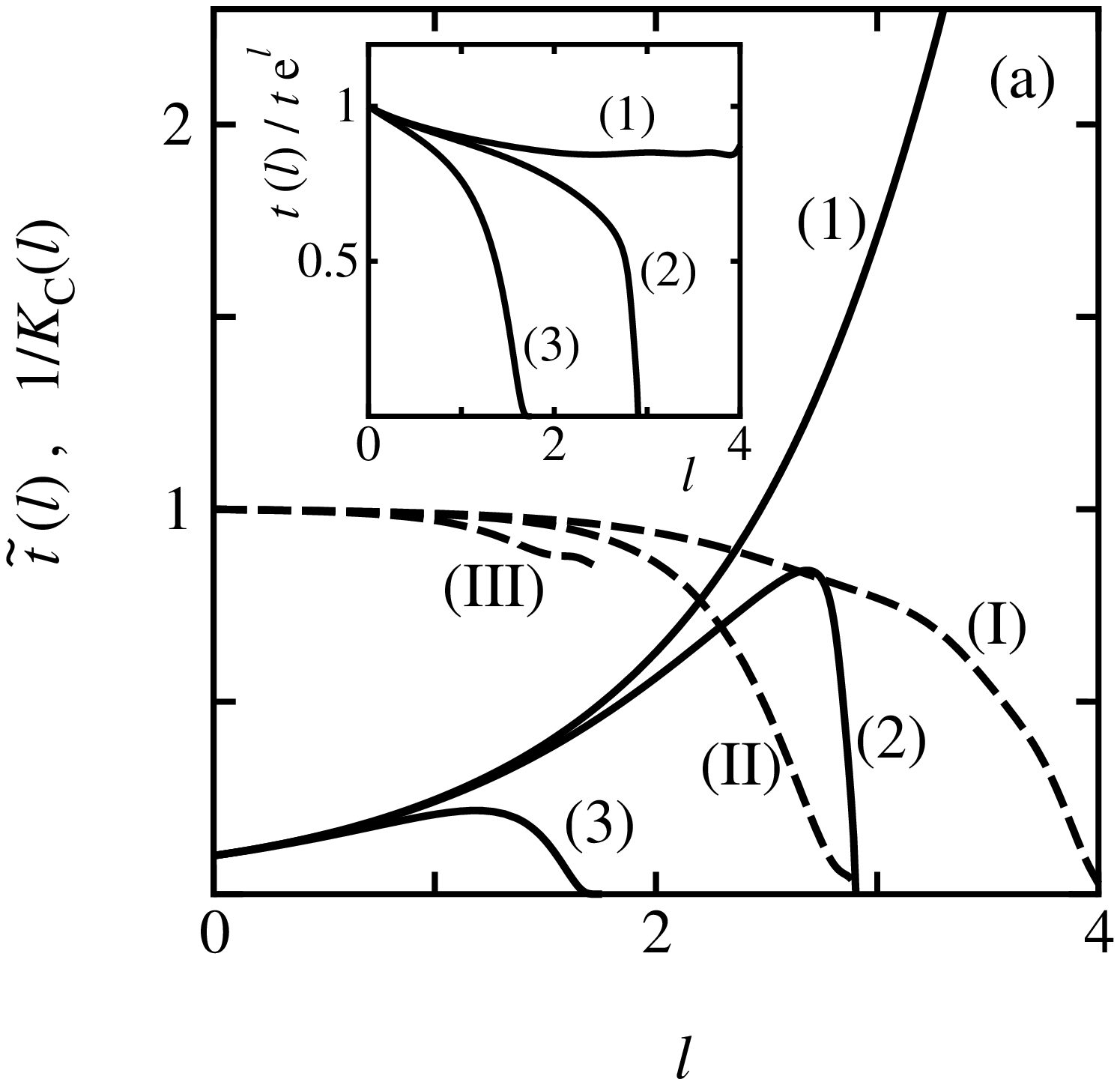}

\epsfxsize=3.5in\epsfbox{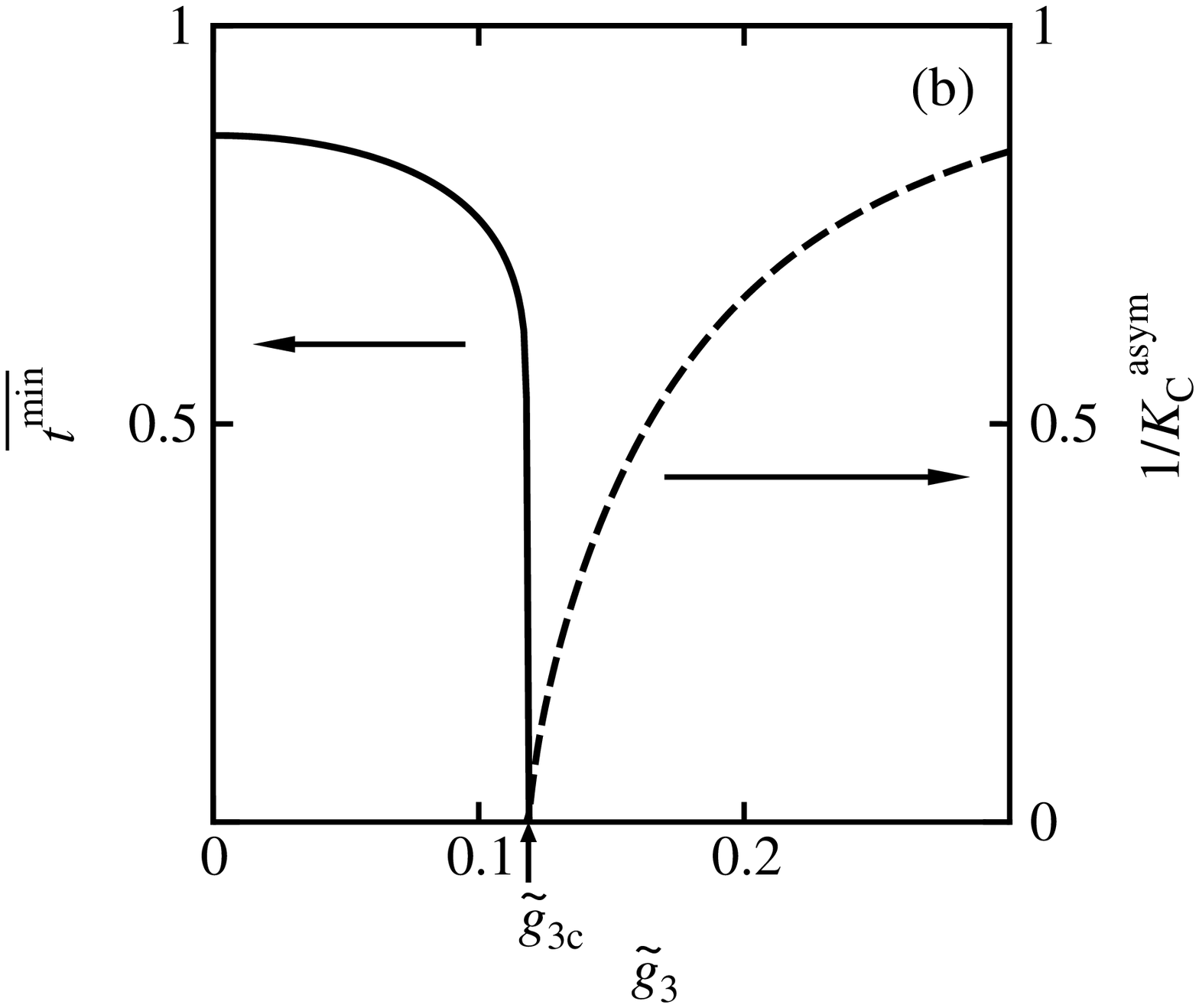}
\end{flushleft}
\caption
{
(a) The $l$-dependences of    
$ \tt (l)$ (solid curves)  
 and   $1/K_{\rm C}(l)$  (dashed curves) 
 for   $\gtu$= 0.05 ((1) and (I)), 
   $\gtu = \gtuc (=0.119)$ ((2) and (II)) and  
     $\gtu = 0.3$ ((3) and (III)), respectively   
 where   $t/\ef=0.1$ and $\gtb = \gtf = 0.4$.
In the inset,  curves (1), (2) and (3) denote 
 $ t (l)/t \e^l$ 
for $\gtu =$ 0.05 (1), $\gtuc$ (2) and 0.3 (3).   
(b)  The $\gtu$-dependences of
  $\overline{t^{\rm min}}$ and
  $1/K_{\rm C}^{\rm asym}$. 
  The quantity $\overline{t^{\rm min}}$ denotes 
  a minimum of  $t(l)/t \e^l$ and 
  the quantity  $1/K_{\rm C}^{\rm asym}$ is  
  the limiting value of $1/K_{\rm C}(l)$.
  The arrow denotes the critical value, $\gtu = \gtuc$, 
  corresponding to a boundary between deconfinement and confinement. 
}
\end{figure}
\noindent
 Eqs.~(\ref{K_theta})-(\ref{G:TL}) 
  obtained with use of  the sharp cutoff in the formulation of
  renormalization group technique.{\cite{Giamarchi_JPF}}
 The fact that  
  $\grcp(l)/\grcm(l)\simeq  \gpcp(l)/\gpcm(l) 
  \simeq \gcpsp(l)/\gcmsp(l) \simeq \gcpsm(l)/\gcmsm(l) 
  \simeq 1/K_{\rm C}(l)$ 
 for the limiting value is consistent with   
   the irrelevance of the interchain hopping. 
 At a  critical value given by   $\gt_3 = \gtuc$,  
  a transition from deconfinement  to confinement
  takes place  where 
  both $\tt (l)$  and   $1/K_{\rm C}(l)$  reduce to zero   
  at the limiting  value of $l$ (curves (2) and (II)). 
 In the inset, 
 the normalized interchain hopping,  
$t (l)/ t \, \e^l$,  is shown 
 for $\gt_3$ = 0.1 (1), 0.119 (2) and 0.3(3)
where
$ t \, \e^l$ denotes the value for non-interacting one. 
 The limiting behavior of curve (1),  
 which remains constant  for  large $l$, 
 indicates deconfinement.
In Fig.~1(b), the $\gt_3$-dependences of 
$1/K_{\rm C}^{\rm asym}$
  and $\overline{t^{\rm min}}$ are shown 
  where $1/K_{\rm C}^{\rm asym}$ is the limiting value  of 
  $K_{\rm C}$.   
  The quantity   $\overline{t^{\rm min}}$ denotes  
 a minimum of   $t(l)/t \e^l$,  which is found  
  with increasing $l$ from zero (e.g. curve (1) in the inset of Fig.~1(a)), 
 and is essentially the same as 
 the limiting value. 
Deconfinement is obtained 
  for finite  $\overline{t^{\rm min}}$  while   
  confinement is found  
  for finite $1/K_{\rm C}^{\rm asym}$.
Both   $\overline{t^{\rm min}}$ and 
 $1/\Kc^{\rm asym}$ are
 reduced to zero at 
  $\gt_3 = \gt_{3c}$,
 which denotes a critical value for deconfinement-confinement
  transition.   
  We note that  
    the Bessel function $J_1(4\tt (l))$ in Eq.~(\ref{dt}) 
  is crucial  to obtain such a transition.
Actually, in r.h.s. of Eq.~(\ref{dt}),  
  the second term  
  becomes negligible for deconfinement
  but the second term becomes larger than the first term 
  for confinement.

In Fig.~2, the corresponding $l$-dependences  
  for  $K_{\rho}(l)$,  $K_{\sigma}(l)$ and   $K_{\rm S}(l)$ 
  are shown by solid curves, dotted curves 
 and dashed curves  respectively
 where  numerical results are shown for $|G_{\nu p, \nu' p'}(l)|<10$. 
   Curves (1), (4) and (7) are for  $\gtu = 0.05$,
    curves (2), (5) and (8) are for $\gtu = \gtuc $ 
    and  curves (3), (6) and (9) are for $\gtu = 0.3$. 
  The quantity $K_{\rho}(l)$ as a function of $l$
 decreases  to zero. A charge gap is formed for 
  $K_{\rho}(l)\simeq K_{\rho}/2$,
 which gives a result consistent with 
that of the Hubbard model.\cite{Woynarovich}
  The transverse spin fluctuation is also 
  suppressed by umklapp scattering because  
 $K_{\rm S}(l)$ with  the fixed $l$ 
 is reduced by  $\gtu$. 
 However the   $\gtu$-dependence  of $K_{\sigma}(l)$ is very small, i.e.,
the $l$-dependence of
   $K_{\sigma}(l)$ is similar to one-dimensional case. 
  Therefore there is no behavior of spin gap 
 for the total spin fluctuation  except for very low  energy. 
We note that, for single chain, 
  $K_{\sigma}(l)$ decreases monotonically to
  $K_{\sigma}(l\to \infty) \to 1$ and that  
  $K_{\rm S}(l)=1$ for all $l$. 
 From these $l$-dependences, it is 
 found that  
  a separation of freedoms of charge and
  spin still exists  at energy corresponding to  formation of 
   the charge gap. 
\begin{figure}[bht]
\begin{flushleft}
\leavevmode
\epsfxsize=3.in\epsfbox{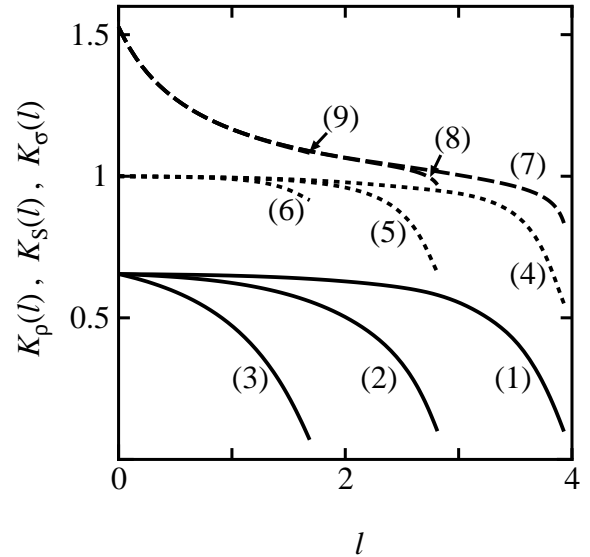}
\end{flushleft}
\caption
{
 The $l$-dependences of 
 $K_{\rho}(l)$, $K_{\rm S}(l)$  and $K_{\sigma}(l)$
 are shown by solid curves, dotted curves
 and dashed curves 
  for $\gtu = 0.05$ ((1), (4), (7)), $\gtuc$ ((2), (5), (8)) and
  $0.3$ ((3), (6), (9)), respectively 
 where parameters are the same as those of Fig.~1(a). 
}
\end{figure}
\noindent

\begin{figure}[t]
\begin{flushleft}
\leavevmode
\epsfxsize=3.in\epsfbox{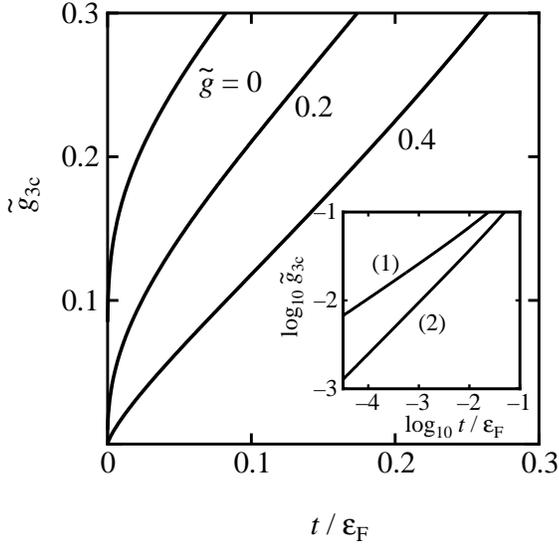}
\end{flushleft}
\caption
{
The critical values, $\gtuc$, are shown as the function of  $t/\ef$ 
 for  $\tilde{g} =0$, $0.2$ and $0.4$,  
 where  confinement (deconfinement) is obtained for 
  $\gtu > \gtuc$ ($\gtu < \gtuc$). 
In the inset, the $\log t/\ef$-$\log \gtuc$ plot 
  is shown for $\gt = 0.2$ (1) and $0.3$ (2).
}
\end{figure}
\noindent

In Fig.~3, the $t$-dependence of $\gtuc$,  
 is shown for
      $\gt_1 = $
  $\gt_2 \equiv \gt = 0$, 
  $0.2$ and $0.4$  
  where  confinement (deconfinement) 
  is obtained  for  $\gtu > \gtuc$ ($\gtu < \gtuc$). 
 The $\gtuc$-dependence of  $t$ for  $\gt = 0$ is expressed as   
\begin{eqnarray} \label{eqn:31}
t/\ef \simeq K_1 \exp[- \pi/ 4 \gt_3] \virg 
 \end{eqnarray}  
  where $K_1 \simeq 1.2$.  
 The  intrachain interaction enhances 
  the confined region. 
 The presence of  $\gt$ leads to a different behavior 
 for $t$ with small $\gtu$.  
From  the inset
 which is calculated  for 
 $\gt =  0.2$ (1) and $0.3$ (2), 
it turns out that 
 the $\gtuc$-dependence of $t$    
 for $\gt \not= 0$  is obtained as 
\begin{eqnarray} \label{eqn:32}
t/\ef \simeq K_2 \left( \gtuc /K_3 \right)^{1/2\gt} \virg 
\label{eqn:power}
 \end{eqnarray}  
  where   $K_2 \simeq 0.2$, $K_3 \simeq  0.2$ and 
  $\gtuc \ll 1$. 
The range of $\gtuc$, in which Eq.~(\ref{eqn:power}) is valid, 
 decreases with 
  decreasing $\gt$, e.g. the upper bound of $\gtuc$ is given by
  $0.15$, $0.04$ and $0$ for $\gt = 0.4$, $0.2$ and $0$ respectively. 
 Coefficients $K_1, K_2$ and $K_3$  
 have been derived numerically 
 since analytical treatment  is very complicated. 
  We note that the magnitude of $\gtuc$ is determined mainly by 
  the balance  between the charge gap 
  created by the umklapp scattering and the energy of interchain hopping
 as is shown later. 
 Coefficients $K_1, K_2$ and $K_3$   
 have been derived numerically 
 since analytical treatment  is very complicated. 
  We note that the magnitude of $\gtuc$ is determined mainly by 
  the balance  between the charge gap 
  created by the umklapp scattering and the energy of interchain hopping
 as is shown later. 
 As $t$ goes to zero, $\gtuc$ reduces to zero and then 
  the interchain hopping  becomes  always relevant 
  in the absence of umklapp scattering 
  within the present choice of intrachain interaction.

\begin{figure}[t]
\begin{flushleft}
\leavevmode
\epsfxsize=3.1in\epsfbox{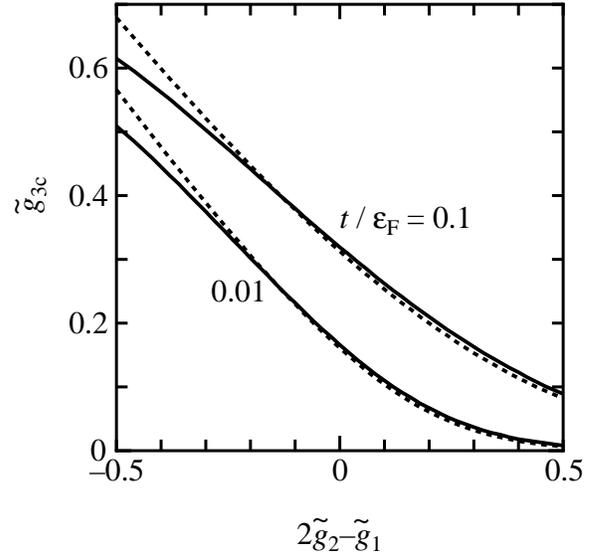}
\end{flushleft}
\caption
{
 The critical values $\gtuc$  as the function of 
 $2\gt_2-\gt_1$ for   $t/\ef =0.1$ and $0.01$
  with the fixed  $\gt_1 = 0.2$. 
 The solid curves denote the boundaries obtained from 
 Eqs.~(2.10)-(2.16) while 
 the dotted curves are obtained in terms of the expanded $K_{\nu}$. 
}
\end{figure}
\noindent

Now we examine the boundary between confinement and deconfinement 
  as the function of $\gt_1$ and $\gt_2$
  with the fixed $t$.
 From the calculation with some choices of $t/\ef = 0.1$, 
  we find it a good approximation  
that  the boundary with fixed $\gtu$ depends only on $2 \gt_2 - \gt_1$.
Therefore,  $\gtuc$ is determined essentially as the function of 
   $2\gt_2 - \gt_1$,  i.e.,   $K_{\rho}$.   
  Such a result originates  in the  fact that $\grcp (l)$-term  
 gives a dominant  contribution 
 and the effect of $g_{\sigma}$ is negligibly small 
 for other coupling constants 
 in  r.h.s. of Eq.~(\ref{dt}).
 In Fig.~4, the quantity  $\gtuc$ as  the function of   $2\gt_2-\gt_1$  
  is shown by the solid curves 
  with choices of $t/\ef = 0.1$ and $0.01$ where $\gt_1 = 0.2$.
The region of $\gtu > \gtuc$ ($\gtu < \gtuc$) 
  corresponds to the confinement (deconfinement). 
 The dotted curves is explained in section IV. 
 With increasing $2\gt_2-\gt_1$ (i.e., decreasing  
  $K_{\rho}$), the region for 
  the confinement is enhanced.

 By use of   response functions for  order parameters 
 (Eqs.~(\ref{O_LSDW})-(\ref{O_SS})),  
 we calculate states
  at finite temperatures where a crossover is shown 
 on the plane of $\gtu$  and normalized temperature, $T/\ef$, 
 (or energy) in Fig.~5.
  The dotted curve denotes the temperature corresponding to  
  the  gap for the total charge,  $\Delta$,  where   
  $\Delta \equiv \ef \exp [-l_{\rm g}]$  
 and   $K_\rho (l_{\rm g}) \equiv K_\rho  /2$. 
 Around the  energy  corresponding to  the charge gap,
 the  total spin  shows a behavior  
similar to one-dimensional case with   the  gapless excitation.  
The solid curve which is obtained from 
  $ \ef \exp [-l_t]$  with  $\tt(l_t) = 1 $ 
denotes the crossover temperature, below which 
 the state reveals
 the property of  two-coupled chains.   
Such a temperature becomes lower than  
  the bare interchain hopping energy, $t/\ef = 0.1$,
  due to the renormalization 
 by the intrachain interaction.\cite{Bourbonnais,Tsuchiizu}
 The dash-dotted line denotes a boundary where 
 the decrease of temperature leads to 
 confinement (deconfinement) 
  for $\gtu > \gtuc$  ($\gtu < \gtuc$).  
 One finds  the following four kinds of regions (I) $\sim$ (IV)  
 which are separated by these boundaries. 
 The dominant state in region (I) is  one-dimensional SDW.
In region (II), interchain hopping 
\begin{figure}[t]
\begin{flushleft}
\leavevmode
\epsfxsize=3.in\epsfbox{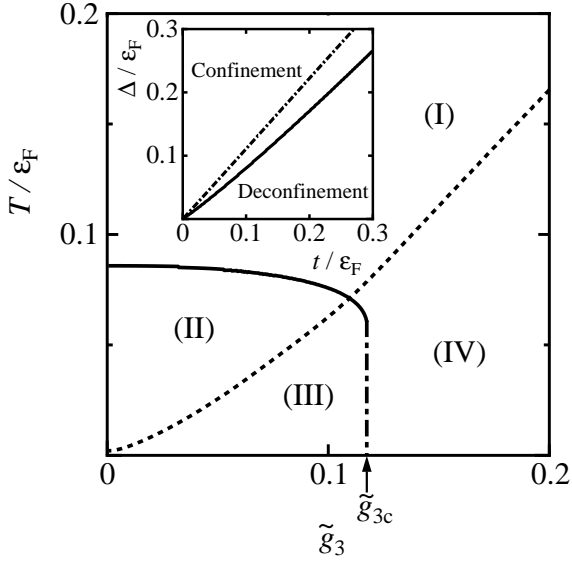}
\end{flushleft}
\caption
{
 A crossover is shown  on the plane of $\gtu$  
  and temperature $T/\ef$ (or energy) where 
 $t/\ef = 0.1$ and  $\gt_1=\gt_2= 0.4$.
 The dotted curve and solid curve denote 
  $\Delta ( \equiv \ef \exp [-l_{\rm g}])$  and $\ef \e^{-l_t}$
  respectively,
  where $K_\rho (l_{\rm g}) = K_\rho /2$ and $\tt(l_t)=1$. 
The dash-dotted line separates the region 
  of confinement from that of deconfinement.
In the inset, the phase diagram of confinement and deconfinement
  is shown   on the plane of $t/\ef$  and  $\Delta/\ef$ where 
 the dash-dotted  (solid) curve  denotes the boundary for
   $\gt =$ 0 (0.4). 
}
\end{figure}
\noindent
 removes the degeneracy of  
  out-of-phase state and in-phase
 state,\cite{Tsuchiizu}  
 but the energy is still high compared with  
  the gaps for the transverse fluctuations, 
 which develop just above $\Delta$ (dotted curve). 
 In this region, one finds a crossover into  
  the out-of-phase SDW state.   
 The present calculation indicates 
 a short range correlation for  the  SCd state  
 in a  certain domain   
 with  small  $\gtu (< 0.001)$  
 and finite temperatures 
  just above the dotted curve.  
 This could retain a trace 
 that the ground state for $\gtu=0$ 
 is given by the SC state. \cite{Fabrizio,Schulz,Balents}
In region (III), 
the gap of the total charge fluctuation develops. 
At very low temperatures, 
 all the fluctuations become gapful due to 
 relevant interchain hopping.
 In this case, 
   the correlation of the SCd state as well as the SDW state 
  decays exponentially.
We note that such a state in the limit of low energy 
 corresponds to the 
  ''C0S0'' phase which has been obtained at half-filled band 
  by Balents and Fisher.\cite{Balents}
 In region (IV), 
the  gap of the total charge 
 is so large that the interchain hopping becomes irrelevant
 leading to the isolated chains\cite{Giamarchi_physica}
 and then absence of other gaps. 
 The state in this region has a resemblance to 
 that of  the half-filled one-dimensional chain. 
\cite{Solyom}
 In the inset, we 
  show a phase diagram of confinement and deconfinement 
 on the plane of $t/\ef$ and $\Delta/\ef$ 
 in the limit of absolute  zero temperature,  
 where   dash-dotted curve (solid curve) corresponds to 
 $\gt =0$ $(0.4)$  in Fig.~3. 
 The region for confinement increases by the increase 
 of intrachain interaction, $\gt$. 
The ratio in the interval region of $0.01 < t/\ef <0.3$ 
  is given   by $\Delta/t \simeq 1.1$ 
 ($0.7 < \Delta /t < 0.9$) 
 for $\gt=0$ $(0.4)$.
\cite{correct}
 These results indicate the fact that the 
 deconfinement-confinement transition is determined 
essentially by the 
competition between the charge gap 
 and  the interchain hopping energy.

\begin{figure}[t]
\begin{flushleft}
\leavevmode
\epsfxsize=3.in\epsfbox{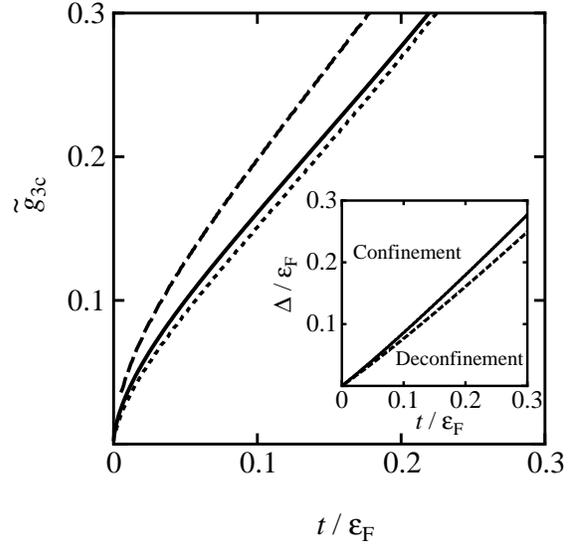}
\end{flushleft}
\caption
{
 The critical value $\gtuc$  as the function of $t/\ef$ 
 for  $\tilde{g} =0.3$. 
The solid curve corresponds to the calculation in Fig.~3 
  and the dotted curve is obtained by use of the expanded $K_{\nu}$. 
The dashed curve is the results for $\gt_1=\gt_2=\gt_4 = 0.3$. 
In the inset, the phase diagram of confinement and deconfinement
  is shown   on the plane of $t/\ef$ and $\Delta/\ef$ 
for $\gt_1=\gt_2= 0.3$ where 
 the solid (dashed) curve is calculated for $\gt_4 = 0$ ($\gt_4 = 0.3$).
}
\end{figure}
\noindent

\section{Discussion}

 In the present paper, the effect of umklapp scattering 
on two-coupled chains has been examined 
 by calculating the boundary for confinement and deconfinement  
as the function  of   umklapp scattering,  interchain hopping 
and intrachain interactions.
 It has been shown that 
 electrons are confined in a single chain 
  when interchain hopping becomes 
 smaller than a critical value of the order of the charge gap or 
 the  umklapp  scattering exceeds a threshold.

  We discuss the validity of our calculation 
 of renormalization group equations given by 
  Eqs.~(\ref{K_theta})-(\ref{dt}).
 Here 
 we calculate these equations  by making use of 
 the expansion,   
  $  K_{\nu}^{\pm 1}(l) = 1 \mp  G_{\nu} (l) + \cdots  $
 ($\nu=\rho$, $\sigma$, ${\rm C}$ and ${\rm S}$), where 
   the initial conditions are   
  $G_{\nu}(0) = g_\nu /2\pi\vf$.
 Since such a method leads to a solution with  the SU(2) symmetry, 
we have done following two kinds of evaluations. 
 One of them is shown in Fig.~4 by  the dotted curve which  denotes 
  $\gt_{3c}$ as a function of 
 intrachain interaction with fixed $t/\ef = 0.1$ and 0.01. 
A good coincidence  between 
  solid curve and dotted curve is obtained for $2\gt_2-\gt_1>0$. 
 The other is shown in Fig.~6 by the dotted curve which 
 denotes  $\gt_{3c}$ as a function of 
 $t/\ef$ for $\gt_1=\gt_2=\gt= 0.3$. 
 There is  a small difference between 
 the result without the expansion of $K_{\nu}$ (solid curve) 
 and  that with the expansion of $K_{\nu}$. 
 Thus,  such a  calculation 
 may be  justified  
  within the present choice of parameters.

 Instead of treating two-coupled chains 
 with the split  Fermi surface,  Khveshchenko and Rice  applied the 
 bosonization method to  two degenerate bands 
 and enlarged the parameter space of the renormalization group to 
 examine  the pair-hopping 
 for the case of the relevant interchain hopping. 
\cite{Rice}   
 We discuss  the effect of  the pair-hopping process    
   on the present result with  the half-filled band.  
  Expressing the   pair-hopping   
  in terms of Eqs.~(\ref{eq:phase_rho})-(\ref{eq:phase_S}), 
      we obtain  non-linear terms consisting of  
      $\theta_{\rho +}$, $\theta_{\sigma +}$,  
       $\theta_{\rm C +}$ and  $\theta_{\rm S +}$, 
     which are also found  in  Eq.~(\ref{eqn:H}) 
      except for  a new term given by 
     $  g_8 \cos ( \sqrt{2}\theta_{\rho +}) 
          \cos (\sqrt{2} \theta_{\sigma +})$.
\cite{Rice}  
When the interchain hopping is larger than the charge gap, 
  the pair-hopping becomes relevant.  
 In such a region,   
 $\tilde{t} (l)$ and $K_{\rho} (l)$ 
 of the present paper exhibit trajectories similar  to those  
of the reference 15 and the  differences in  magnitudes are  small, 
 although  the  $g_8$-term gives rise to a visible enhancement of    
  the spin gap obtained  by $K_{\sigma}(l)$.   
 It is noted that 
 the  interchain hopping is not renormalized directly by 
  the $g_8$-term 
  since the r.h.s. of  Eq.~(\ref{dt}) is determined 
 only  by  terms  including $\theta_{\rm C +}$. 
  When the interchain hopping is smaller than the charge gap, 
   the pair-hopping becomes irrelevant due to confinement, e.g., 
     $g_8 \simeq 0$ at the energy of the charge  gap.  
 Therefore it is considered that  
 the effect of the pair-hopping  is small  for 
  the boundary between  confinement and deconfinement.

 Now we examine   the effect of the forward 
  scattering with the same branch, 
 where the coupling constant is given by $g_4$.
 The Hamiltonian corresponding to $g_4$
  is expressed as    
\begin{eqnarray}
{\cal H}_{\rm int}^{g_{4}} 
&\equiv&
    \frac{g_{4}}{2L} \sum_{p,\sigma,i} \sum_{k_1,k_2,q}
\nonumber \\ &&
    a_{k_1,p,\sigma,i}^\dagger a_{k_2,p,-\sigma,i}^\dagger
    a_{k_2+q,p,-\sigma,i} a_{k_1-q,p,\sigma,i}   
      \virg 
\end{eqnarray}
  which has two kinds  effects.  
 One of them appears in   $K_{\rho}$, $K_{\sigma}$, 
  $v_{\rho}$ and $v_{\sigma}$ which are  written as  
$K_\rho = 
     [ \{1 -\gt_{\rho}/(1+\gt_4) \}
      / \{1 +\gt_{\rho}/(1+\gt_4) \}]^{1/2}$, 
$K_\sigma = 
     [ \{1 -\gt_{\sigma}/(1-\gt_4) \}
      / \{1 +\gt_{\sigma}/(1-\gt_4) \}]^{1/2}$, 
$v_\rho = \vf [(1+\gt_4)^2-\gt_{\rho}^2]^{1/2}$ and
$v_\sigma = \vf [(1-\gt_4)^2-\gt_{\sigma}^2]^{1/2}$ with
$\gt_{\nu} = g_{\nu}/2 \pi \vf$. 
 Another is the  nonlinear terms given by  
\begineq
\begin{eqnarray}
\frac{1}{2\pi^2 \alpha^2} \int \d x
\left\{ 
           g_{{\rm C}+, {\rm C}-}
           \cos \left( \sqrt{2}\theta_{{\rm C}+} 
                        - \frac{4t}{\vf}x \right) 
           \cos \sqrt{2} \theta_{{\rm C}-} 
   +    \, g_{{\rm S}+, {\rm S}-}
           \cos \sqrt{2} \theta_{{\rm S}+}
            \cos \sqrt{2} \theta_{{\rm S}-}  
\right\}
                                   \virg 
\label{H:g4}
\end{eqnarray}
\endeq
  where $g_{{\rm C}+, {\rm C}-} = - g_{{\rm S}+, {\rm S}-} = g_{4}$. 
 Equation (\ref{H:g4}) leads to additional terms 
  to renormalization group equations 
 for $\tt (l)$ and $K_{\rm C}(l)$, which are of the order of 
 $o(\tt^3)$ and $o(\tt^4)$ respectively. 
There is also renormalization for $G_{\rm C+,C-}$ and $G_{\rm S+.S-}$.
 However we found that, in the present calculation,
 the latter effect  given by Eq.~(\ref{H:g4}) is negligibly small. 
 On the other hand, in the former case, 
 there is a  noticeable  effect of $g_4$, which 
 comes from the variation of $K_{\rho}$. 
 Note that  $K_{\rho} (< 1)$  
 increases with increasing $\gt_4$, 
 since  the effect of  $g_4$  is equivalent to 
 replacing   $\gt_{\rho}$   by   
 $\gt_{\rho}/(1+\gt_4)$ in $K_{\rho}$.   
 The increase of $K_{\rho} (<1)$ 
  reduces 
  the renormalization of umklapp scattering, $G_{\rho +, {\rm C} +}$. 
 Then one needs larger $\gtu$ to obtain  the confinement.  
 An example with $\gt = 0.3$   is shown in Fig.~6
 where  $\gtuc$ including  the $\gt_{4}$-term (dashed curve) is 
 compared with   that with $\gt_4=0$ (solid curve). 
   The quantity 
   $\gt_{3c}$ is increased  by   $g_{4}$, i.e.,   
the suppression of    the region of the confinement 
 on the plane of $t/\ef$ and $\gtu$.  
 In the inset, a  phase diagram 
 of confinement and deconfinement is shown on the plane of $t/\ef$ 
 and $\Delta/\ef$  for  both  $\gt_4=0$ (solid curve)  
 and $\gt_4 \not= 0$ (dashed curve). 
 The fact that   the critical value of $\Delta$ 
  is reduced  by $g_4$ 
  is understood as follows. 
 The  magnitude of $K_{\rho}(0) $  for $\gt_4 \not= 0$ 
 is larger than that  for $\gt_4 = 0$, 
 although the difference between these two cases is very small as for 
 r.h.s. of Eq.~(\ref{K_theta}). 
 Since larger $K_{\rho}(0)$  leads to slower decrease of $K_{\rho} (l)$,  
  one obtains  smaller  gap, $\Delta$, in the presence of $g_4$.  
 Thus,  it is found that   $g_4$ reduces the effects of both 
  intrachain interaction 
 and  umklapp scattering.

 Based on  the present calculation, 
 we comment on electronic  states for TMTSF and TMTTF salts, both of which 
show the correlation gap  above the SDW state.   
 Such a gap is possible for a choice of 
 $T/\ef \simeq 10^{-2}$ in Fig.~5    
 where   $\ef \simeq 10^{3} K$ and $T \simeq 10 K$.\cite{Jerome}  
 The fact that the  plasma edge perpendicular to chains 
 is present  for TMTSF salt  (absent for TMTTF salt)\cite{Gruner_prepri} 
 suggests the relevance (irrelevance) of the interchain hopping. 
These behaviors of interchain hopping are found when   
  the umklapp scattering satisfies  
    $\gtu \lsim \gtuc$ for TMTSF salt and 
 $\gtu \gsim \gtuc$ for the TMTTF salt.  
  Thus  
 the existence of dimerization, which increases $\gtu$,  
   is crucial to understand 
  property of these  salts above the  SDW state.

 For the metallic state,  the  above organic conductors may be regarded as a doped system rather  half-filling    
 since one-particle hopping between chains leads to small deviation from commensurate one.\cite{Giamarchi_physica}  
 In this case, the metallic state is expected 
   with increasing  the doping rate 
 as is shown in a one-dimensional case.\cite{Mori} 
 Actually a crossover from confinement to deconfinement has been obtained 
 in the presence of doping  even for two-coupled chains.\cite{Suzumura_ICSM98}
It will be of interest to study such an effect of doping on 
 many coupled chains.

\acknowledgments

 The authors thank 
  G. Gr\"uner and H. Yoshioka for useful discussions.  
 This work was partially supported by  a Grant-in-Aid 
 for Scientific  Research  from the Ministry of Education, 
Science, Sports and Culture, (No.09640429) Japan.

\end{multicols}
\appendix
\begin{multicols}{2}\columnseprule 0pt\narrowtext\noindent

\section{Phase Representation of ${\cal H}$ and  Order Parameters}

By making use of $a_{k,p,\sigma,i}
  =\{ (-1)^i c_{k,p,\sigma,+}+c_{k,p,\sigma,-} \}$ ($i = 1$, 2) and 
  $\psi_{p,\sigma,\mu}(x) = L^{-1/2} 
  \sum_k \e^{ikx} c_{k,p,\sigma,\mu}$, 
  Eq.~(\ref{H0}) is rewritten in terms of $\psi_{p,\sigma,\mu}$.
The terms for interactions are divided into two parts,
  which consist of the scattering between electrons with same $\mu$
  and the scattering between electrons with opposite $\mu$.
Since the former part is treated in the way similar to the kinetic 
  energy, we examine the latter part which is defined as 
  $\cal{H}_{\rm int}$.
By defining $\psi_{p,\sigma,\mu}'$ as
  $\psi_{p,\sigma,\mu}'(x) = 
  (1/\sqrt{2\pi \al}) \exp \left( ipk_{{\rm F}\mu}x 
   + i\Theta _{p,\sigma,\mu} \right) $,
  ${\cal H}_{\rm int}$  is written as 
\begineq
\begin{eqnarray}
{\cal H}_{\rm int}
&=& \frac{1}{4} \sum_{p,\sigma,\mu} \int \d x \,\,
 \biggl[
      g_{1} \,\,
      \psi_{p,\sigma,\mu}'^\dagger \, \psi_{-p,\sigma,\mu}'^\dagger \,
      \psi_{p,\sigma,-\mu}' \, \psi_{-p,\sigma,-\mu}' 
      \,\, 
      \e^{i(\sigma \mu) 2\pi\N_a}
\nonumber \\ && \nonumber \\ 
&&{}+ 
     g_{1} \,\,
      \psi_{p,\sigma,\mu}'^\dagger \, \psi_{-p,-\sigma,\mu}'^\dagger \,
      \psi_{p,-\sigma,-\mu}' \, \psi_{-p,\sigma,-\mu}'   
      \,\, 
      \e^{i(p\sigma) 2\pi \N_d + i\pi}
\nonumber \\ && \nonumber \\ 
&&{}  + 
      g_{1} \,\,
      \psi_{p,\sigma,\mu}'^\dagger \, \psi_{-p,\sigma,-\mu}'^\dagger \,
      \psi_{p,\sigma,-\mu}' \, \psi_{-p,\sigma,\mu}'
      \,\,  
      \e^ {i (p\sigma \mu )2\pi \N_b + i\pi }
\nonumber \\ && \nonumber \\ 
&&{}+ 
     g_{1} \,\,
      \psi_{p,\sigma,\mu}'^\dagger \, \psi_{-p,-\sigma,-\mu}'^\dagger \,
      \psi_{p,-\sigma,-\mu}' \, \psi_{-p,\sigma,\mu}'   
      \,\, 
      \e^{i(p\sigma) 2\pi\N_d}
\nonumber \\ && \nonumber \\ 
&&{}+  
     g_{1} \,\,
      \psi_{p,\sigma,\mu}'^\dagger \, \psi_{-p,-\sigma,\mu}'^\dagger \,
      \psi_{p,-\sigma,\mu}' \, \psi_{-p,\sigma,\mu}'  
      \,\,  
      \e^{-i (p\sigma) 2\pi[\N_b + \mu \N_d] + i\pi}
\nonumber \\ && \nonumber \\ 
&&{}+  
     g_{1} \,\,
      \psi_{p,\sigma,\mu}'^\dagger \, \psi_{-p,-\sigma,-\mu}'^\dagger \,
      \psi_{p,-\sigma,\mu}' \, \psi_{-p,\sigma,-\mu}'  
      \,\,  
      \e^{-i(p\sigma) 2\pi  [\N_b + p\mu \N_c] + i\pi}
\nonumber \\ && \nonumber \\ 
&&{}+ 
     g_2 \,\,
      \psi_{p,\sigma,\mu}'^\dagger \, \psi_{-p,\sigma,-\mu}'^\dagger \,
      \psi_{-p,\sigma,\mu}' \, \psi_{p,\sigma,-\mu}'  
      \,\, 
      \e^{i(p\sigma \mu )2\pi  \N_b} 
\nonumber \\ && \nonumber \\ &&{}+ 
     g_2 \,\,
      \psi_{p,\sigma,\mu}'^\dagger \, \psi_{-p,-\sigma,-\mu}'^\dagger \,
      \psi_{-p,-\sigma,\mu}' \, \psi_{p,\sigma,-\mu}'   
      \,\,  
      \e^{i (\sigma \mu ) 2\pi \N_a+ i\pi}
\nonumber \\&& \nonumber \\ 
&&{}+ 
     g_2 \,\,
      \psi_{p,\sigma,\mu}'^\dagger \, \psi_{-p,\sigma,\mu}'^\dagger \,
      \psi_{-p,\sigma,-\mu}' \, \psi_{p,\sigma,-\mu}' 
      \,\,  
      \e^{i ( \sigma \mu) 2\pi  \N_a+ i\pi}
\nonumber \\ && \nonumber \\ &&{}+ 
     g_2 \,\,
      \psi_{p,\sigma,\mu}'^\dagger \, \psi_{-p,-\sigma,\mu}'^\dagger \,
      \psi_{-p,-\sigma,-\mu}' \, \psi_{p,\sigma,-\mu}'  
      \,\,  
      \e^{i ( p\sigma \mu) 2\pi \N_b}
 \nonumber \\&& \nonumber \\ 
&&{}+ 
     g_{3} \,\, 
      \e^{ip4\kf x}\,
      \psi_{p,\sigma,\mu}'^\dagger \, \psi_{p,-\sigma,\mu}'^\dagger \,
      \psi_{-p,-\sigma,\mu}' \, \psi_{-p,\sigma,\mu}' 
      \,\,  
      \e^{i p 4\pi [\N_a+ \delta_{\mu,+}\N_b-\delta_{\mu,-}\N_d]}
\nonumber \\ && \nonumber \\ &&{}+ 
     g_{3} \,\, 
      \e^{ip4\kf x}\,
      \psi_{p,\sigma,\mu}'^\dagger \, \psi_{p,-\sigma,-\mu}'^\dagger \,
      \psi_{-p,-\sigma,-\mu}' \, \psi_{-p,\sigma,\mu}'  
      \,\, 
      \e^{i ( p\sigma \mu) 2\pi  \N_b+ i\pi}
\nonumber \\&& \nonumber \\ 
&&{}+ 
     g_{3} \,\, 
      \e^{ip4\kf x}\,
      \psi_{p,\sigma,\mu}'^\dagger \, \psi_{p,-\sigma,-\mu}'^\dagger \,
      \psi_{-p,-\sigma,\mu}' \, \psi_{-p,\sigma,-\mu}'  
      \,\, 
      \e^{i ( \sigma \mu) 2\pi  \N_a} 
\nonumber \\ && \nonumber \\ &&{}+
     g_{3} \,\, 
      \e^{ip4\kf x}\,
      \psi_{p,\sigma,\mu}'^\dagger \, \psi_{p,-\sigma,\mu}'^\dagger \,
      \psi_{-p,-\sigma,-\mu}' \, \psi_{-p,\sigma,-\mu}' 
      \,\, 
      \e^{i p 2\pi[ \delta_{p\mu, +}\{4\N_a-\N_7-\N_8\} 
                  +\delta_{p\mu, -}\{\N_3+\N_4\} ]} 
\biggr]  \virg
\label{eqn:ap-Hint}
\end{eqnarray} 
\endeq
  where $\N_a = [(\N_1+\N_2)+(\N_3+\N_4)+(\N_5+\N_6)+(\N_7+\N_8)] /4$,
  $\N_b =[\{(\N_1+\N_2)+(\N_3+\N_4)\}-\{(\N_5+\N_6)+(\N_7+\N_8)\}]/4 $,
$\N_c = [\{(\N_1+\N_2)-(\N_3+\N_4)\}+\{(\N_5+\N_6)-(\N_7+\N_8)\}]/4$   
 and
$\N_d = [\{(\N_1+\N_2)-(\N_3+\N_4)\}-\{(\N_5+\N_6)-(\N_7+\N_8)\}]/4$.
By defining $N_i$ as the eigen value of $\N_i$,  
  it is found that  the factor in Eq.~(\ref{eqn:ap-Hint}) 
  commutes with the Hamiltonian, Eq.~(\ref{H0}), when  
$N_a$, $N_b$, $N_c$ and $N_d$
 are integers. 
Such a  choice of  Hilbert space leads to   
  negative sign for 2, 3, 5, 6, 8, 9  and 12-th terms 
  in Eq.~(\ref{eqn:ap-Hint}).
By expressing  $\psi_{p,\sigma,\mu}'(x)$ 
  in terms of the phase variables,  
  we obtain the non-linear terms in Eq.~(\ref{phase_Hamiltonian}).

Next we examine order parameters.
 For order parameter,  
 $O_{{\rm LSDW}_{\para,{\rm out}}}$, which is expressed as
$O_{{\rm LSDW}_{\para,{\rm out}}}
= - \sum_{\sigma ,\mu} \sigma \, 
         \psi_{+,\sigma,\mu}^\dagger  \, 
         \psi_{-,\sigma,-\mu}  
= - ( 
         \psi_{1}^\dagger  \, \psi_{7} 
+ \psi_{3}^\dagger  \, \psi_{5}  )
+ (
         \psi_{2}^\dagger  \, \psi_{8}  
+          \psi_{4}^\dagger  \, \psi_{6} )  $, 
 we evaluate correlation function  given by
$\langle O_{{\rm LSDW}_{\para,{\rm out}}}(x) $ $
     O_{{\rm LSDW}_{\para,{\rm out}}}^\dagger (0) \rangle$. 
By noting that a typical term of this correlation function 
  is rewritten as 
\begin{eqnarray}
&&
\lan
\left(  \psi_{1}^\dagger(x)  \, \psi_{7}(x) \right)
\left( \psi_{2}^\dagger (0) \, \psi_{8}(0) \right)^\dagger
\ran 
\nonumber \\ && \nonumber \\ 
&=&
\lan
 \psi_{1}'^\dagger(x)  \, \psi_{7}'(x) \, 
  \psi_{8}'^\dagger(0)  \, \psi_{2}'(0)\, 
  \e^{-i2\pi [\N_b+\N_c]+i\pi}
\ran  \nonumber \\&& \nonumber \\ 
&=&-\lan
\left( \psi_{1}'^\dagger(x)  \, \psi_{7}'(x) \right)
\left( \psi_{2}'^\dagger(0)  \, \psi_{8}'(0) \right)^\dagger
\ran ,
\end{eqnarray} 
  the correlation function is rewritten as,
\begin{eqnarray}
&&
\lan
\left[
-\left(     \psi_{1}^\dagger  \, \psi_{7} + \psi_{3}^\dagger  \, \psi_{5}  \right)
+ \left(    \psi_{2}^\dagger  \, \psi_{8}  +  \psi_{4}^\dagger  \, \psi_{6} \right)
\right]
\right. \nonumber \\ &&  \nonumber \\ && {} \times \left.
\left[
-\left(     \psi_{1}^\dagger  \, \psi_{7} + \psi_{3}^\dagger  \, \psi_{5}  \right)
+ \left(    \psi_{2}^\dagger  \, \psi_{8}  +  \psi_{4}^\dagger  \, \psi_{6} \right)
\right]^\dagger
\ran \nonumber \\&& \nonumber \\ 
&=&
\lan
\left[
\left(     \psi_{1}'^\dagger  \, \psi_{7}'+ \psi_{3}'^\dagger  \, \psi_{5}' \right)
+ \left(\psi_{2}'^\dagger \, \psi_{8}'  +  \psi_{4}'^\dagger  \, \psi_{6}' \right)
\right]
\right. \nonumber \\&& \nonumber \\  && {} \times \left.
\left[
\left(  \psi_{1}'^\dagger  \, \psi_{7}' + \psi_{3}'^\dagger \,\psi_{5}'  \right)
+ \left(\psi_{2}'^\dagger  \, \psi_{8}'  +  \psi_{4}'^\dagger  \, \psi_{6}'\right)
\right]^\dagger
\ran .  
\end{eqnarray}
Therefore $O_{{\rm LSDW}_{\para,{\rm out}}}$ 
 in the response function can be expressed as 
\begin{eqnarray}
O_{{\rm LSDW}_{\para,{\rm out}}}
&=& - \sum_{\sigma ,\mu} \sigma \, 
         \psi_{+,\sigma,\mu}^\dagger  \, 
         \psi_{-,\sigma,-\mu} 
\nonumber \\
&\to& \sum_{\sigma ,\mu}
         \psi_{+,\sigma,\mu}'^\dagger \, 
         \psi_{-,\sigma,-\mu}'   \virg
\end{eqnarray}
 which leads to 
  Eq.~(\ref{O_LSDW}) with 
 $\psi_{p,\sigma,\mu}'(x) = 
(1/\sqrt{2\pi \al})$ $ \exp \left( ipk_{{\rm F}\mu}x 
 + i\Theta _{p,\sigma,\mu} \right) $.
The other order parameters are obtained in a  similar way.

\section{Derivation of renormalization group equations}

We evaluate  response functions 
 by use of the renormalization 
  group method.\cite{Giamarchi_PRB,Tsuchiizu}
By treating the nonlinear terms 
  in Eq.~(\ref{phase_Hamiltonian})
  as the perturbation, the response function for 
  $\theta_{{\rm S}+}$ field is calculated
  up to the third order as
\begineq
\begin{eqnarray}            \label{A1}
&&
\lan T_\tau 
   \exp \left[ (i/\sqrt{2})\, \theta_{{\rm S}+} (x_1,\tau_1)\right] 
\,
   \exp \left[-(i/\sqrt{2})\, \theta_{{\rm S}+} (x_2,\tau_2)\right] 
  \ran 
  \nonumber \\&& \nonumber \\ 
&=&  \e^{-(\Ks/2)U (r_1\f-r_2\f)}    \nonumber \\&& \nonumber \\ 
&& + \sum_{\nu = \rho, \sigma} 
    \frac{1}{(4\pi)^2 \tvn^2} \sum_{\epsilon = \pm 1} 
     \int \frac{\d ^2 r_3^\nu}{\alpha^2}
          \frac{\d ^2 r_4^\nu}{\alpha^2}
     \e^{-(\Ks/2) U (r_1\f-r_2\f)}
     \e^{-2\Knu    U (r_3^\nu-r_4^\nu)}\nonumber \\&& \nonumber \\ 
&& \hspace{1cm} \times \biggl\{
    \gnsp^2 \e^{-2\Ks    U (r_3\f-r_4\f)}
     \left(
       \e^{
               \epsilon \Ks \{
    U(r_1\f-r_3\f)-U(r_1\f-r_4\f)-U(r_2\f-r_3\f)+U(r_2\f-r_4\f)
               \}
            } -1
     \right) \nonumber \\&& \nonumber \\ 
&& \hspace{2.5cm} + \gnsm^2 \e^{-(2/\Ks)    U (r_3\f-r_4\f)}
     \left(
       \e^{ 
             i  \epsilon \{
    U(r_1\f-r_3\f)-U\f(r_1\f-r_4\f)-U(r_2\f-r_3\f)+U(r_2\f-r_4\f)
               \}
            } -1
     \right) \biggr\} \nonumber \\&& \nonumber \\ 
&& + \frac{1}{(4\pi)^2}  \sum_{\epsilon = \pm 1} 
     \int \frac{\d ^2 r_3\f}{\alpha^2}\frac{\d ^2 r_4\f}{\alpha^2}
     \e^{-(\Ks/2) U (r_1\f-r_2\f)} \nonumber \\&& \nonumber \\ 
&& \hspace{.5cm} \times \biggl[  \left\{
            \gcpsp^2 \e^{-2\Kc U(r_3\f-r_4\f)} 
                       \cos 2q_0 (x_3-x_4)
          + \gcmsp^2 \e^{-(2/\Kc)U(r_3\f-r_4\f)}
                       \right\}   \nonumber \\&& \nonumber \\ 
&& \hspace{3.5cm} \times 
          \e^{-2\Ks    U (r_3\f-r_4\f)} 
     \left(
       \e^{ 
               \epsilon \Ks \{
    U(r_1\f-r_3\f)-U(r_1\f-r_4\f)-U(r_2\f-r_3\f)+U(r_2\f-r_4\f)
               \}
           } -1
     \right) \nonumber \\&& \nonumber \\ 
&& \hspace{2cm} + \left\{
            \gcpsm^2 \e^{-2\Kc U(r_3\f-r_4\f)} 
                       \cos 2q_0 (x_3-x_4)
          + \gcmsm^2 \e^{-(2/\Kc)U(r_3\f-r_4\f)}
                       \right\}     
      \nonumber \\&& \nonumber \\ 
&& \hspace{4.5cm} \times 
      \e^{-(2/\Ks) U (r_3\f-r_4\f)}
     \left(
       \e^{ 
              i \epsilon \{
    U(r_1\f-r_3\f)-U(r_1\f-r_4\f)-U(r_2\f-r_3\f)+U(r_2\f-r_4\f)
               \}
            } -1
     \right)  \biggr] \nonumber \\&& \nonumber \\ 
&& - \sum_{\nu = \rho, \sigma}
    \frac{4}{(4\pi)^3 \tvn^2} \sum_{\epsilon = \pm 1} 
     \int \frac{\d ^2 r_3^{\rm F}}{\alpha^2} 
          \frac{\d ^2 r_4^\nu}{\alpha^2}
          \frac{\d ^2 r_5^\nu}{\alpha^2}
     \e^{-(\Ks/2)U(r_1\f-r_2\f)} 
     \e^{-2\Knu U(r_4^\nu-r_5^\nu)} 
          \nonumber \\&& \nonumber \\ 
&& \hspace{1cm} \times \biggl[ \Bigl\{
          \gncp \gnsp \gcpsp \e^{-2\Kc U(r_3\f-r_5\f)}  
                        \cos 2q_0 (x_3-x_5)
\nonumber \\ && \nonumber \\ && \hspace{2.5cm}
        + \gncm \gnsp \gcmsp \e^{-(2/\Kc)U(r_3\f-r_5\f)}
                   \Bigr\} 
      \,  \e^{-2\Ks U(r_3\f-r_4\f)}     \nonumber \\&& \nonumber \\ 
&& \hspace{3cm} \times 
     \left(
       \e^{
               \epsilon \Ks \{
       U(r_1\f-r_3\f)-U(r_1\f-r_4\f)-U(r_2\f-r_3\f)+U(r_2\f-r_4\f)
               \}
           } -1
     \right) \nonumber \\&& \nonumber \\ 
&& \hspace{2cm} + \Bigl\{
          \gncp \gnsm \gcpsm \e^{-2\Kc U(r_3\f-r_5\f)}  
                        \cos 2q_0 (x_3-x_5)
\nonumber \\ && \hspace{3.5cm}
        + \gncm \gnsm \gcmsm \e^{-(2/\Kc) U(r_3\f-r_5\f)}
                   \Bigr\}  
             \, \e^{-(2/\Ks) U(r_3\f-r_4\f)}  \nonumber \\&& \nonumber \\ 
&& \hspace{4cm} \times  
     \left(
       \e^{
               i \epsilon \{
       U(r_1\f-r_3\f)-U(r_1\f-r_4\f)-U(r_2\f-r_3\f)+U(r_2\f-r_4\f)
               \}
           } -1
     \right) \biggr]  +  \cdots \virg
\end{eqnarray}
\endeq
  where  
  $U(r_i^\nu - r_j^\nu) 
  = \ln \left[\sqrt{(x_i - x_j)^2 
         + v_\nu^2 (\tau_i - \tau_j)^2}/\alpha\right]$ 
  for $i,j = 1, 2, 3, 4$,  
   $\d^2r^\nu = v_\nu \, \d x \, \d \tau$
 ($\nu = \rho$, $\sigma$ and ${\rm F}$) and $q_0 \equiv 2t/\vf$.
In order to obtain scaling equations 
  of the coupling constants up to the second order,
  we need to  calculate 
   response functions up to the third order 
  for the nonlinear terms. Note that 
 these  terms do not exist  in 
  one-dimensional case.\cite{Giamarchi_JPF} 
By putting $r_5=r_4+r$ and $r_5=r_3+r$, and expanding near $r=0$,
 we obtain the following renormalization  in terms of 
 effective quantities,
\begineq
\begin{eqnarray}
\Ks^{\rm eff} &=& \Ks
-    \frac{1}{2} \sum_{\nu = \rho, \sigma} 
                  \gnsp^2 \Ks^2 \int \frac{\d r\f}{\alpha}
                  \biggl( \frac{r\f}{\alpha} \biggr)
                   ^{3-2\Knu-2\Ks}
+    \frac{1}{2} \sum_{\nu = \rho, \sigma} 
                  \gnsm^2 \int \frac{\d r\f}{\alpha}
                  \biggl( \frac{r\f}{\alpha} \biggr)
                   ^{3-2\Knu-2/\Ks}
\nonumber \\&& \nonumber \\ 
&& \hspace{1cm} 
-\frac{1}{2} \gcpsp^2 \Ks^2 \int \frac{\d r\f}{\alpha}
                  \biggl( \frac{r\f}{\alpha} \biggr)
                   ^{3-2\Kc-2\Ks}
                  J_0(2q_0r\f)    
\nonumber \\&& \nonumber \\ 
&& \hspace{1cm}
+ \frac{1}{2} \gcpsm^2 \int \frac{\d r\f}{\alpha}
                  \biggl( \frac{r\f}{\alpha} \biggr)
                   ^{3-2\Kc-2/\Ks}
                  J_0(2q_0r\f)    
\nonumber \\&& \nonumber \\ 
&& \hspace{1cm}
-\frac{1}{2} \gcmsp^2 \Ks^2 \int \frac{\d r\f}{\alpha}
                  \biggl( \frac{r\f}{\alpha} \biggr)
                   ^{3-2/\Kc-2\Ks}
+\frac{1}{2} \gcmsm^2 \int \frac{\d r\f}{\alpha}
                  \biggl( \frac{r\f}{\alpha} \biggr)
                   ^{3-2/\Kc-2/\Ks}     ,
\label{eqn:K_phitilde}
\\ && \nonumber \\ 
{\gns^{\rm eff}}^2 &=& \gns^2 
    -2 \gcps \gncp \gns \int \frac{\d r\f}{\alpha} 
           \biggl( \frac{r\f}{\alpha} \biggr)^{1-2\Kc}
           J_0(2q_0 r\f)
\nonumber \\&& \nonumber \\  && \hspace{1.3cm} 
    -2 \gcms \gncm \gns \int \frac{\d r\f}{\alpha} 
           \biggl( \frac{r\f}{\alpha} \biggr)^{1-2/\Kc}\virg 
\label{eqn:G_k} 
\\ && \nonumber \\ 
{\gcs^{\rm eff}}^2 &=& \gcs^2 
    -\sum_{\nu = \rho, \sigma}
     \frac{2}{\tvn} \gcs \gnc \gnsd \int \frac{\d r^\nu}{\alpha} 
           \biggl( \frac{r^\nu}{\alpha} \biggr)^{1-2\Knu}
\virg 
\label{eqn:G_a} 
\end{eqnarray}
\endeq
where $r^\nu = (x^2 + (v_\nu \tau)^2)^{1/2}$,
  $\tilde{v}_\nu = v_\nu / \vf$ and
  $p$, $p' = \pm$.
The second and third terms of r.h.s. 
  in Eqs.~(\ref{eqn:G_k})-(\ref{eqn:G_a}) are obtained by exponentiating
  the third order terms of Eq.~(\ref{A1}).
For the transformation given by  
  $\alpha \to \alpha' = \alpha \e^{\d l}$, \cite{Giamarchi_JPF}
these quantities are scaled as
\begin{eqnarray}
\Ks^{\rm eff}(K ',G',q_0',\alpha') 
  &=& \Ks^{\rm eff}(K,G,q_0,\alpha),
\\ \nonumber \\
G_{\nu p, \nu' p'}^{\rm eff}(K ',G',q_0',\alpha') 
  &=& G_{\nu p, \nu' p'}^{\rm eff}(K,G,q_0,\alpha)    
\nonumber \\ &&\nonumber \\&& {}\times 
  \left( \alpha'/\alpha \right)^{\gamma_{\nu p, \nu' p'}},
\end{eqnarray}
  where $K '$, $G'$ and $q'_0$ denote renormalized quantities. 
The exponent $\gamma_{\nu p, \nu' p'}$ is given by
$\gamma_{\nu p, \nu' p'} = 2-K_{\nu}^p - K_{\nu'}^{p'}$.
 By applying this infinitesimal transform to 
  Eqs.~(\ref{eqn:K_phitilde}),
  (\ref{eqn:G_k})
  and  (\ref{eqn:G_a}),
  we obtain 
 renormalization equations given by 
  Eqs.~(\ref{K_phitilde}), (\ref{G:Phi})
  and (\ref{G:TL}),  respectively.
In a similar way,
  the  renormalization group equation for 
  $K_\nu(l)$ ($\nu = \rho$, $\sigma$ and ${\rm C}$)
  is calculated from the response function given by 
  $\langle T_\tau \e^{ (i/\sqrt{2})\theta_{\nu \pm} (x_1,\tau_1)} 
               \e^{-(i/\sqrt{2})\theta_{\nu \pm} (x_2,\tau_2)} \rangle$ 
  and the equations for $G_{\nu +, {\rm C} \pm}(l)$ and 
  $G_{\nu +, {\rm S} \pm}(l)$ ($\nu = \rho$ and $\sigma$)
 are calculated from the response function for
  $\theta_{\nu \pm}$ field.\cite{T2}
We note that, in case of $t=0$, 
 these equations become equal to the  one-dimensional 
  equations.\cite{Solyom} 

The renormalization equation for $\tt(l)$ is 
 derived by evaluating  the  difference of 
 the density between two bands ($\mu = \pm$), which is given by 
\end{multicols}\widetext
\begin{eqnarray}
\Delta n  &\equiv&  2 ( k_{{\rm F}+} - k_{{\rm F}-}) \alpha 
          + 2 \frac{T}{L} \int \d x \, \d \tau 
            \lan 
            \widetilde{k}_{{\rm F}+} - \widetilde{k}_{{\rm F}-}
            \ran  \alpha    
\nonumber \\ &&\nonumber \\ 
&=& - 2 q_0 \alpha + \sqrt{2} \frac{T}{L} \alpha
             \int \d x \, \d \tau \, 
             \lan \partial_x \, \theta_{{\rm C} +} (x,\tau) \ran
\nonumber \\ &&\nonumber \\ 
&=& -2 q_0 \alpha 
   + \sum_{\nu = \rho, \sigma}
     \frac{4}{\alpha} \gncp \Kc \frac{T}{L}
     \int \d x \, \d \tau
     \lan x \sin \left(\sqrt{2}\theta_{{\rm C}+}-2q_0x \right)
            \cos \sqrt{2}\theta_{\nu +}  \ran
\nonumber \\&&\nonumber \\  && \hspace{1.5cm}
   + \sum_{p = \pm}
     \frac{4}{\alpha} \gcps \Kc \frac{T}{L} 
     \int \d x \, \d \tau
     \lan x \sin \left(\sqrt{2}\theta_{{\rm C}+} -2q_0x \right)
            \cos \sqrt{2}\theta_{{\rm S} p}  \ran
\nonumber \\ &&\nonumber \\ 
&=& -2 q_0 \alpha 
  + \sum_{\nu = \rho, \sigma}
    \gncp^2 \Kc \int \frac{\d r\f}{\alpha} 
    \biggl(\frac{r\f}{\alpha}\biggr)^{2-2\Kc-2\Knu}
    J_1(2q_0r\f)
\nonumber \\ &&\nonumber \\ && \hspace{1.5cm}
  + \sum_{p = \pm}
    \gcps^2 \Kc \int \frac{\d r\f}{\alpha} 
    \biggl(\frac{r\f}{\alpha}\biggr)^{2-2\Kc-2\Ks^p}     J_1(2q_0r\f)
  + \ldots  \virg   \label{eqn:dn}
\end{eqnarray}
where $4( \widetilde{k}_{{\rm F}+} 
  - \widetilde{k}_{{\rm F}-})/2\pi 
  \equiv \sum_{p,\sigma, \mu} \mu \,
  \psi^\dagger_{p, \sigma, \mu} \psi_{p, \sigma, \mu}$.
The assumption of scaling invariance of 
 Eq.~(\ref{eqn:dn}) with respect to 
infinitesimal transformation $\alpha' = \alpha \e^{\d l}$ 
  leads to  Eq.~(\ref{dt}).

The response functions   are calculated 
 in terms of the solutions of Eqs.~(\ref{K_theta})-(\ref{dt}). 
 The response function for $O_{{\rm SS}_{\perp,{\rm in}}}$,
  which is defined by 
$R_{{\rm SS}_{\perp,{\rm in}}}(x_1-x_2,\tau_1-\tau_2) 
\equiv 
\langle T_\tau O_{{\rm SS}_{\perp,{\rm in}}}(x_1,\tau_1)
            O_{{\rm SS}_{\perp,{\rm in}}}^\dagger (x_2,\tau_2) \rangle$,
  is calculated by writing 
  $R_{{\rm SS}_{\perp,{\rm in}}}(x,\tau) 
    \equiv R_{{\rm SS}_{\perp,{\rm in}}}^{(0)}(x,\tau) \, 
      F_{{\rm SS}_{\perp,{\rm in}}}(x,\tau)$
  where  
$R_{{\rm SS}_{\perp,{\rm in}}}^{(0)}(x,\tau) 
  = \bigl(\alpha/\sqrt{x^2+v_\rho ^2 \tau^2}\bigr)^{1/2\Krho}$
$\bigl(\alpha/\sqrt{x^2+v_\sigma   ^2 \tau^2}\bigr)^{\Ksigma/2} $
$    \bigl(\alpha/\sqrt{x^2+\vf^2 \tau^2}\bigr)
    ^{(1/\Kc + \Ks)/2}$. 
By assuming the scaling relation 
$F_{{\rm SS}_{\perp,{\rm in}}} (r, \al (l), K_\nu(l),G(l)) 
= I_{{\rm SS}_{\perp,{\rm in}}}(\d l,K_\nu(l),G(l)) 
  \cdot F_{{\rm SS}_{\perp,{\rm in}}}
  (r,\al (l+\d l),K_\nu(l+\d l),G(l+\d l))$, 
  the multiplicative factor $I_{{\rm SS}_{\perp,{\rm in}}}$
  is written as, 
\begin{eqnarray}
I_{{\rm SS}_{\perp,{\rm in}}} 
&=& 
\exp \biggl[ - \gpcm \d l + \gpsp \d l  + \gcmsp \d l  \nonumber \\&&\nonumber \\ 
&& \hspace{.6cm}
        + \frac{1}{4\tvt^2}\biggl\{- \grcp^2 J_0 (2 q_0 \alpha )  
                  -  \grcm^2 
                  -  \grsp^2 
                  -  \grsm^2 
          \biggr\}  U(r_1^{\rho}-r_2^{\rho}) \, 
                       \d l   
          \nonumber \\&&\nonumber \\ 
&& \hspace{.6cm} 
        + \frac{1}{4\tvp^2} \Ksigma^2 
              \biggl\{ \gpcp^2 J_0(2q_0\al)
                 +  \gpcm^2 
                 +  \gpsp^2 
                 +  \gpsm^2 
          \biggr\}  U(r_1^{\sigma}-r_2^{\sigma}) \, \d l  
\nonumber \\&&\nonumber \\ 
&& \hspace{.6cm}
        + \frac{1}{4} 
           \biggl\{ \sum_{\nu = \rho, \sigma} \biggl( 
                 -  \gncp^2 J_0(2q_0\al)
                 +  \gncm^2    /\Kc^2
                 + \gnsp^2 \, \Ks^2
                 - \gnsm^2      \biggr)
                   \nonumber \\&&\nonumber \\ 
&& \hspace{2cm}
                + \gcpsp^2 \left( -1 + \Ks^2 \right)   J_0(2q_0\al)
                 + \gcpsm^2 \left( - 2 \right)  J_0(2q_0\al)
           \nonumber \\&&\nonumber \\ 
&& \hspace{2cm}
                 + \gcmsp^2
                   \left( 1/\Kc^2 + \Ks^2 \right)
                 + \gcmsm^2
                   \left( 1/\Kc^2 - 1 \right) 
          \biggr\} \,  U(r_1\f-r_2\f) \, \d l 
\biggr] \virg 
\end{eqnarray}
 which leads to   $F_{{\rm SS}_{\perp,{\rm in}}}$ 
 expressed as
\begin{eqnarray}
F_{{\rm SS}_{\perp,{\rm in}}} (r,K,G) 
=\exp \left[ 
        \sum_{l=0}^{\ln (r/\al)} \ln 
        \left[ I_{{\rm SS}_{\perp,{\rm in}}}(\d l,K(l),G(l)) \right]
     \right].
\end{eqnarray}
 We note that  terms  including the second order of the coupling 
 constants are rewritten in a simple form. 
 For example, one obtains 
\begin{eqnarray}
&& \frac{1}{4} \int \d l 
\biggl\{ \sum_{\nu = \rho, \sigma} \biggl(
                 - \gncp^2  J_0 (2 q_0 \alpha )
                 + \gncm^2 /\Kc^2
                 + \gnsp^2 \, \Ks^2 
                 - \gnsm^2   \biggr)
            +\gcpsp^2 \left( -1 + \Ks^2 \right)      J_0(2q_0\al)
\nonumber \\&&\nonumber \\  && \hspace{1.5cm}
                 + \gcpsm^2 \left( - 2 \right)  J_0(2q_0\al)
          \Biggr. 
                 + \gcmsp^2
                   \left( 1/\Kc^2 + \Ks^2 \right)
                 + \gcmsm^2
                   \left( 1/\Kc^2 - 1 \right) 
          \biggr\}  \ln \left[ \frac{r}{\al (l)} \right]  \nonumber \\&&\nonumber \\ 
&&= \frac{1}{2} \int \d l \left\{
        \frac{1}{\Kc^2(l)} 
        \frac{\d \Kc(l)}{\d l}  
      - \frac{\d \Ks(l)}{\d l} \right\}   
  \ln \left[ \frac{r}{\al (l)} \right] 
\nonumber \\&&\nonumber \\ 
&&= \frac{1}{2} 
     \left\{ \frac{1}{\Kc(0)} +  \Ks(0) \right\}
      \ln \left(\frac{r}{\al}\right)
    - \int_{0}^{\ln(r/\al)} \d l \frac{1}{2}
      \left\{ \frac{1}{\Kc(l)}  
                 +  \Ks(l) \right\} \virg
\end{eqnarray}
  where 
  $\alpha (l) = \alpha \e^{l}$. 
 Thus, the normalized response function, 
  $\overline{R}_{{\rm SS}_{\perp,{\rm in}}} (x,\tau)$
  ($\equiv R_{{\rm SS}_{\perp,{\rm in}}} (x,\tau) \cdot 2(\pi \alpha)^2$),
  is expressed as,
\begin{eqnarray}
\overline{R}_{{\rm SS}_{\perp,{\rm in}}} (x,\tau)  
&=&  \exp \left[
     \int_0^{\ln \sqrt{x^2+(v_\rho \tau)^2}/\alpha  } 
          \d l 
        \left\{
           - \frac{1}{2} \frac{1}{\Krho (l)}
        \right\}   \right]   
\nonumber \\&&\nonumber \\  &&  \times
\exp \left[
     \int_0^{\ln \sqrt{x^2+(v_\sigma \tau)^2}/\alpha } 
          \d l 
        \left\{ - \frac{1}{2} \Ksigma(l) - \frac{1}{\tvp}\gpcm (l) +\frac{1}{\tvp} \gpsp (l) \right\}
     \right]
\nonumber \\&&\nonumber \\  && \times
\exp \left[
     \int_0^{\ln \sqrt{x^2+(\vf \tau)^2}/\alpha } 
          \d l 
        \left\{ - \frac{1}{2} \left(
               \frac{1}{\Kc (l)} + \Ks(l) \right)
              + \gcmsp (l)
        \right\}
  \right]  \virg  \label{eqn:app-final}
\end{eqnarray}
 which leads to Eq.~(\ref{R-SS}). 
 Other response functions are obtained in a similar way.
In deriving Eqs.~(\ref{R-LSDW})-(\ref{R-SS}), 
  we replaced  $v_\rho$ and $v_\sigma$ by $\vf$, 
  which may cause a slight deviation of the numerical factor.

\begin{multicols}{2}
\columnseprule 0pt
\narrowtext

\end{multicols}


\begin{thebibliography}{99}
\bibitem{correct}
The second term of the r.h.s. of Eq.~(\ref{dt}) is corrected to 
 be twice as large as that of  the previous work 
  (reference 26). 
  The resultant  region of confinement  
  becomes  slightly large.
\end{thebibliography}
\end{document}